\documentclass[twocolumn,showpacs,preprintnumbers,prc,floatfix]{revtex4}
\usepackage{amsmath}
\usepackage{amssymb}
\usepackage{graphicx}
\usepackage{dcolumn}
\usepackage{epsfig,color}
\usepackage{bm}
\usepackage{longtable}

\begin{document}
\title{Strong decays of baryons and missing resonances}
\author{R. Bijker} 
\affiliation{Instituto de Ciencias Nucleares, Universidad Nacional Aut\'onoma de M\'exico, AP 70-543, 04510 M\'exico DF, M\'exico}
\author{J. Ferretti}
\affiliation{Instituto de Ciencias Nucleares, Universidad Nacional Aut\'onoma de M\'exico, AP 70-543, 04510 M\'exico DF, M\'exico}
\affiliation{Dipartimento di Fisica and INFN, ``Sapienza" Universit\`a di Roma, P.le Aldo Moro 5, I-00185 Roma, Italy \footnote{Present address}}
\author{G. Galat\`a}
\affiliation{Instituto de Ciencias Nucleares, Universidad Nacional Aut\'onoma de M\'exico, 04510 M\'exico DF, M\'exico}
\author{H. Garc{\'{\i}}a-Tecocoatzi}
\affiliation{Instituto de Ciencias Nucleares, Universidad Nacional Aut\'onoma de M\'exico, 04510 M\'exico DF, M\'exico}
\affiliation{INFN, Sezione di Genova and Dipartimento di Fisica,Universit\`a di Genova, via Dodecaneso 33, 16146 Genova, Italy} 
\author{E. Santopinto}
\thanks{Corresponding author: santopinto@ge.infn.it}
\affiliation{INFN, Sezione di Genova, via Dodecaneso 33, 16146 Genova, Italy}

\begin{abstract}
We provide results for the open-flavor strong decays of strange and non-strange baryons into a baryon-vector/pseudoscalar meson pair. The decay amplitudes are computed in the $^3P_0$ pair-creation model, where $s\bar{s}$ pair-creation suppression is included for the first time in the baryon sector, in combination with the U(7) and hypercentral models. 
The effects of this $s\bar{s}$ suppression mechanism cannot be re-absorbed in a redefinition of the model parameters or in a different choice of the $^3P_0$ model vertex factor. 
Our results for the decay amplitudes are compared with the existing experimental data and previous $^3P_0$ and elementary meson emission model calculations. 
In this respect, we show that distinct quark models differ in the number of missing resonances they predict and also in the quantum numbers of states. Therefore, future experimental results will be important in order to disentangle different models of baryon structure. 
Finally, in the appendices, we provide some details of our calculations, including the derivation of all relevant flavor couplings with strangeness-suppression.  
This derivation may be helpful to calculate the open-flavor decay amplitudes starting from other models of baryons. 
\end{abstract}

\pacs{12.39.-x, 13.30.Eg, 14.20.Gk, 14.20.Jn}
\maketitle

\section{Introduction}
It is well known that the baryon spectrum is very rich and much more complex than the meson spectrum.
Nevertheless, at the moment, the number of known light-quark mesons is much larger than the number of known baryon resonances \cite{Nakamura:2010zzi}.
This problem has to do with the difficulty of identifying those high-lying baryon resonances that are only weakly coupled to the $N \pi$ channel \cite{Capstick:1992uc,Capstick:1992th} and thus cannot be seen in elastic $N \pi$ scattering experiments. 
Since the experimental observations of baryon resonances mainly come from reactions in which the pion is present either in the incoming channel, such as $N \pi \rightarrow N \pi$, or in the outgoing one, such as $N \gamma \rightarrow N \pi$, it would not be surprising if some baryon resonances, very weakly coupled to the single pion, were missing from experimental results. 
These baryons may decay mainly into two pion channels ($N \pi \pi$) or into channels such as $N \eta$, $N \eta'$, $N \omega$ and $K^+ \Lambda$, where the final-state meson is different from the pion \cite{Capstick:1992th}. Although interesting results were provided by CB-ELSA \cite{Crede:2003ax}, TAPS \cite{Krusche:1995nv}, GRAAL \cite{Renard:2000iv}, SAPHIR \cite{Tran:1998qw} and CLAS \cite{Dugger:2002ft}, theoretical calculations of strong, electromagnetic and weak decays of baryons may still help experimentalists in their search for those resonances that are still unknown.

The QCD mechanism behind the OZI-allowed strong decays \cite{Okubo:1963fa} is still not clear. 
For this reason, several phenomenological models have been developed in order to carry out this type of study, including pair-creation models \cite{Micu,LeYaouanc,Eichten:1978tg,Alcock:1983gb,Dosch:1986dp,Kokoski:1985is,Roberts:1992,Ackleh:1996yt}, elementary meson emission models \cite{Becchi,Faiman:1968js,Koniuk:1979vy,Godfrey:1985xj,Sartor:1986qr,Bijker:1996tr} and effective Lagrangian approaches (for example, see Ref. \cite{Colangelo:2012xi}).
Attempts at modeling strong decays within the quark model (QM) formalism date from Micu's suggestion \cite{Micu} that hadron decays proceed through $q \bar q$ pair production with vacuum quantum numbers, i.e. $J^{PC} = 0^{++}$.
Since the $q \bar q$ pair corresponds to a $^3P_0$ quark-antiquark state, this model is known as the $^3P_0$ pair-creation model \cite{Micu,LeYaouanc,Roberts:1992}. 
A few years after its introduction, Le Yaouanc $et$ $al.$ used the $^3P_0$ model to compute meson and baryon open-flavor strong decays \cite{LeYaouanc} and also evaluated the strong decay widths of $\psi(3770)$, $\psi(4040)$ and $\psi(4415)$ charmonium states \cite{LeYaouanc02}.
The $^3P_0$ model, extensively applied to the decays of light mesons and baryons \cite{Blundell:1995ev}, has recently been applied to heavy meson strong decays in the charmonium \cite{Barnes:2005pb,Ferretti:2013faa,Ferretti:2014xqa}, bottomonium \cite{bottomonium,Ferretti:2013vua}, open-charm \cite{Close:2005se,Segovia:2012cd,Ferretti:2015rsa,Godfrey:2015dva} and open-bottom \cite{Ferretti:2015rsa} sectors. 
In the 90's, Capstick and Roberts calculated the $N \pi$ and the strange decays of non-strange baryons \cite{Capstick:1992th} by using relativized { quark} model { wave functions} for baryons and mesons. The baryon and meson spectra were predicted within the relativized QMs of Refs. \cite{Godfrey:1985xj,Capstick:1986bm}. 
It is also worthwhile to cite Ref. \cite{Xiao:2013xi}, where the authors computed the open-flavor strong decays of $\Xi$ baryons up to $N = 2$ shell in a chiral quark model.

In this paper, we present our results for the open-flavor strong decay widths of light baryons (i.e. made up of $u, d, s$ valence quarks) into a baryon-pseudoscalar meson pair and a baryon-vector meson pair. The widths are computed within a modified version of the $^3P_0$ pair-creation model where, for the first time in the baryon case, a strange quark pair suppression mechanism has been taken into account, analogous to what was done in the meson sector to suppress unphysical heavy quark pair-creation \cite{Kalashnikova:2005ui,Ferretti:2013faa,Ferretti:2013vua,Ferretti:2014xqa,bottomonium}. The effects of this mechanism, which breaks the SU(3) symmetry, cannot be re-absorbed in a redefinition of the model parameters or in a different choice of the $^3P_0$ model vertex factor.  

In the next section, we briefly mention the models for the baryon spectrum and structure that we have used for the calculation of the strong decays: the  $U(7)$ algebraic model \cite{Bijker:1994yr-first,Bijker:1994yr}, by Bijker, Iachello and Leviatan, and the hypercentral model (hQM) \cite{pl}, developed by Giannini and Santopinto.
In Section~\ref{3P0 model}, we review the $^{3}P_{0}$ model for the two-body decay of a baryon into a baryon and a meson, including a discussion of the phase space factor. 
The results for the strong decays are presented in Section~\ref{Strong decay widths}. Finally, we present a summary and conclusions. Some details of the calculation of $^{3}P_{0}$ matrix elements are presented in the appendices. 
Of particular interest is Appendix~\ref{Ap2}, where the flavor couplings with $SU(3)$ breaking induced by the presence of the strange pair suppression mechanism have been derived explicitly for the first time. 

\section{Strange and non-strange baryon spectra}
\subsection{$U(7)$ algebraic model}
\label{U(7) algebraic model}
The baryon spectrum is computed by means of algebraic methods introduced by Bijker, Iachello and Leviatan \cite{Bijker:1994yr-first,Bijker:1994yr}. The algebraic structure of the model consists of combining the symmetry of the internal spin-flavor-color part, $SU_{\rm sf}(6) \otimes SU_{\rm c}(3)$, with that of the spatial part, $U(7)$ into 
\begin{equation} 
U(7) \otimes SU_{\rm sf}(6) \otimes SU_{\rm c}(3) ~.
\end{equation}
The $U(7)$ model was introduced \cite{Bijker:1994yr-first} to describe the relative motion of the three constituent parts of the baryon. The general idea is to introduce a so-called spectrum generating algebra $U(k+1)$ for quantum systems characterized by $k$ degrees of freedom. For baryons, there are the $k=6$ relevant degrees of freedom of the two relative Jacobi vectors 
\begin{eqnarray}
	\vec{\rho} &=& \frac{1}{\sqrt{2}} (\vec{r}_1-\vec{r}_2) ~,
\nonumber\\
	\vec{\lambda} &=& \frac{1}{\sqrt{6}} (\vec{r}_1+\vec{r}_2-2\vec{r}_3) ~,
\label{jacobi}
\end{eqnarray}
and their canonically conjugate momenta, $\vec{p}_{\rho} = (\vec{p}_1-\vec{p}_2)/\sqrt{2}$ and 
$\vec{p}_{\lambda} = (\vec{p}_1+\vec{p}_2-2\vec{p}_3)/\sqrt{6}$.
The $U(7)$ model is based on a bosonic quantization which consists of introducing two vector boson operators, $b^{\dagger}_{\rho}$ and $b^{\dagger}_{\lambda}$,  
associated to the Jacobi vectors, and an additional auxiliary scalar boson, $s^{\dagger}$.  
The scalar boson does not represent an independent degree of freedom, but is added 
under the restriction that the total number of bosons $N$ is conserved. The model space consists 
of harmonic oscillator shells with $n=0,1,\ldots,N$.  

The baryon mass formula is written as the sum of three terms
\begin{equation}
	\hat{M}^2 = M_0^2 + \hat{M}^2_{\rm space} + \hat{M}^2_{\rm sf}  \mbox{ },  
\end{equation}
where $M_0^2$ is a constant, $\hat{M}^2_{\rm space}$ is a function of the spatial degrees of freedom and $\hat{M}^2_{\rm sf}$ depends on the internal degrees of freedom. The spin-flavor part is treated in the same way as in Ref.~\cite{Bijker:1994yr} in terms of a generalized G\"ursey-Radicati formula \cite{Gursey:1992dc}, which in turn is a generalization of the Gell-Mann-Okubo mass formula \cite{Ne'eman,Gell-Mann:1962xb} 
\begin{eqnarray}
\hat{M}^2_{\rm sf} &=& a \left( \hat{C}_2(SU_{\rm sf}(6))-45\right) + b \left(\hat{C}_2(SU_{\rm f}(3))-9\right) 
\nonumber\\
&& + c \left(\hat{C}_2(SU_{\rm s}(2))-\frac{3}{4}\right) + d \left(\hat{C}_1(U_{\rm Y}(1))-1\right)  
\nonumber\\ 
&& + e \left(\hat{C}_2(U_{\rm Y}(1))-1\right)+ f \left(\hat{C}_2(SU_{\rm I}(2))-\frac{3}{4} \right)  \mbox{ }.
\nonumber\\
\mbox{}
\end{eqnarray}
The operators $\hat{C}_1(G)$ and $\hat{C}_2(G)$ correspond to the linear and quadratic Casimir invariants of the relevant groups for the internal degrees of freedom. The values of the coefficients $M_0^2$, $a$, $b$, $c$, $d$, $e$ and $f$ are taken from \cite{Bijker:1994yr}.

Since the space-spin-flavor wave function is symmetric under permutation group $S_3$ of the three identical constituents, the permutation symmetry of the spatial wave function has to be the same as that of the spin-flavor part. Thus, the spatial part of the mass operator $\hat{M}^2_{\rm space}$ has to be invariant under the $S_3$ permutation symmetry. 
The dependence of the mass spectrum on the spatial degrees of freedom is given by:
\begin{equation}
\hat{M}^2_{\rm space} = \hat{M}^2_{\rm vib} + \hat{M}^2_{\rm rot}  \mbox{ }.  \label{eqn:U(7)02}
\end{equation}
In Refs.~\cite{Bijker:1994yr,Bijker:1996tr}, strong decays of baryons were studied in the collective string model, which is a special case of $U(7)$ in which the radial excitations are interpreted as rotations and vibrations of an oblate top, in combination with the elementary emission model for the strong decays.
Here, we do the same, but calculate the decays in the $^3P_0$ pair-creation model \cite{Micu,LeYaouanc,Roberts:1992}.
The rotational part of the operator (\ref{eqn:U(7)02}) is written in the same form as in \cite{Bijker:1994yr}
\begin{equation}
	\hat{M}^2_{\rm rot} = \alpha \sqrt{\hat{L} \cdot \hat{L} + \frac{1}{4}} \mbox{ } ,
\end{equation}
with eigenvalues 
\begin{equation}
M^2_{\rm rot} = \alpha \left(L + \frac{1}{2} \right) \mbox{ } \mbox{.}  \label{eqn:U(7)05}
\end{equation}
In this way one gets linear Regge trajectories with a slope $\alpha$, as required by the phenomenology 
\cite{Johnson:1975sg}. 
The spectrum of the vibrational part is given by \cite{Bijker:1994yr}
\begin{eqnarray}
\hat{M}^2_{\rm vib} = M^2_{\rm vib} &=& \kappa_1 \, v_1 + \kappa_2 \, v_2 ~ \mbox{ },  
\label{eqn:U(7)03}
\end{eqnarray}
where $v_1=n_u$ and $v_2=n_v+n_w$ are the vibrational quantum numbers, corresponding to the symmetric stretching vibration along the direction of the strings (breathing mode), and two degenerate bending vibrations of the strings. The spectrum consists of a series of vibrational excitations characterized by the labels $(v_1,v_2)$, and a tower of rotational excitations built on top of each vibration. $\kappa_1$ and $\kappa_2$ are free parameters fitted to the data.

The spectra calculated for the non-strange and strange baryons are shown in Tables~\ref{tab:nuc}--\ref{tab:omega}, second/third column. 
The baryon wave functions are denoted in the standard form as  
\begin{equation}
\left| \, ^{2S+1}\mbox{dim}\{SU_f(3)\}_J \,
[\mbox{dim}\{SU_{sf}(6)\},L_i^P] \, \right> ~, \label{wf}
\end{equation}
where $S$ and $J$ are the spin and total angular momentum $\vec{J}=\vec{L}+\vec{S}$~.
As an example, in this notation the nucleon and delta wave functions are given by $\left| \, ^{2}8_{1/2} \, [56,0_1^+] \, \right>$ and $\left| \, ^{4}10_{3/2} \, [56,0_1^+] \, \right>$, respectively. 

\subsection{Hypercentral model}
\label{hQM}
The starting point of the hQM is the assumption that quark interaction is hypercentral, i.e. it only depends on the hyperradius $x$ \cite{pl,chin},
\begin{equation}
  	V_{3q}(\vec{\rho},\vec{\lambda}) = V(x) \mbox{ },
\end{equation}
with $x = \sqrt{\vec{\rho}^2 + \vec{\lambda}^2}$ \cite{baf}. Thus, the spatial part of the three-quark wave function, $\psi_{space}$, is factorized as
\begin{equation}
	\psi_{space} = \psi_{3q}(\vec{\rho},\vec{\lambda}) = \psi_{\gamma \nu}(x){Y}_{[{\gamma}]l_{\rho}l_{\lambda}}({\Omega}_{\rho},{\Omega}_{\lambda},\xi) \mbox{ },
\label{psi}
\end{equation}
where the hyperradial wave function, $\psi_{\gamma \nu}(x)$, is labeled by the grand angular quantum number $\gamma$ and the number of nodes $\nu$. ${Y}_{[{\gamma}]l_{\rho}l_{\lambda}}({\Omega}_{\rho},{\Omega}_{\lambda},\xi)$ are the hyperspherical harmonics, with angles ${\Omega}_{\rho}=({\theta}_{\rho},{\phi}_{\rho})$, ${\Omega}_{\lambda}=({\theta}_{\lambda},{\phi}_{\lambda})$ and hyperangle $\xi = \arctan {\frac{\rho}{\lambda}}$ \cite{baf}.
The dynamics is contained in $\psi_{\gamma \nu}(x)$, which is a solution of the hyperradial equation
\begin{equation}
	\begin{array}{l}
	[\frac{{d}^2}{dx^2}+\frac{5}{x}~\frac{d}{dx}-\frac{\gamma(\gamma+4)}{x^2}] \psi_{\gamma \nu}(x) \\
	\hspace{1cm} = \mbox{ } - 2m~[E-V_{3q}(x)]~~\psi_{\gamma \nu}(x)  \mbox{ }.
	\end{array}
\label{hyrad}
\end{equation}
In the hQM, the quark interaction has the form \cite{pl,chin}
\begin{equation}
	V(x) = -\frac{\tau}{x} + \alpha x \mbox{ },
	\label{h_pot}
\end{equation}
where $\tau$ and $\alpha$ are free parameters fitted to the reproduction of the experimental data. Eq. (\ref{h_pot}) can be seen as the hypercentral approximation of a Cornell-type quark interaction \cite{Eichten:1978tg}, whose form can be reproduced by Lattice QCD calculations \cite{LQCD}.
Now, to introduce splittings within the SU(6) multiplets, an SU(6)-breaking term must be added. In the case of the hQM, such violation of the SU(6) symmetry is provided by the hyperfine
interaction \cite{deru,ik}. The complete hQM Hamiltonian is then \cite{pl,chin}
\begin{equation}
	H_{\rm hQM} = 3m + \frac{\vec{p}_\rho^{~2}}{2m} + \frac{\vec{p}_\lambda^{~2}}{2m}-\frac{\tau}{x} +
	\alpha x + H_{\rm hyp} \mbox{ },
	\label{H_hCQM}
\end{equation}
where $\vec{p}_\rho$ and $\vec{p}_\lambda$ are the momenta conjugated to the Jacobi coordinates $\vec \rho$ and $\vec \lambda$.
In addition to $\tau$ and $\alpha$, there are two more free parameters in the hQM, the constituent quark mass, $m$, and the strength of the hyperfine interaction. 
The former is taken, as usual, as $1/3$ of the nucleon mass. The latter, as in the case of $\tau$ and $\alpha$, is fitted in \cite{pl} to the reproduction of the *** and **** resonances reported in the PDG \cite{Nakamura:2010zzi}. The hQM has an approximate $O(7)$ symmetry \cite{pl}. 

\subsection*{Extension to strange baryons}
In Ref. \cite{hQM-strange}, the hQM was extended to strange baryons. Specifically, the authors considered a Hamiltonian whose SU(6) invariant part is the same as in the Hypercentral Model~\cite{pl}, while the SU(6) symmetry is broken by a G\"ursey-Radicati-inspired interaction \cite{Gursey:1992dc}.   

The complete Hamiltonian is given by
\begin{equation}\label{eq:tot}
H = H_0 + H_{\rm GR}~,
\end{equation}
where 
\begin{equation}
  H_0=3m+\frac{\mbox{\boldmath $p$}_{\lambda}^2}{2m}+\frac{\mbox{\boldmath $p$}_\rho^2}{2m}+V(x)  
\end{equation}  
and
\begin{equation}
	V(x)= -\frac{\tau}{x}~+~\alpha x  \mbox{ }.
\end{equation}
The G\"ursey-Radicati-inspired interaction, $H_{\mathrm{GR}}$, which describes the splittings within the SU(6) baryon multiplets \cite{hQM-strange}, is written in terms of Casimir operators as,
\begin{eqnarray}
  \label{eq:grfull}
  H_{\mathrm{GR}} &=& A \, \hat{C}_2(SU_{\rm sf}(6)) + B \, \hat{C}_2(SU_{\rm f}(3)) \nonumber\\
       &&+C \, \hat{C}_2(SU_{\rm s}(2)) + D \,\hat{C}_1(U_{\rm Y}(1)) \nonumber\\
       &&+E \left( \hat{C}_2(SU_{\rm I}(2)) - \frac{1}{4}\hat{C}^2_1(U_{\rm Y}(1))\right)  \mbox{ },
\end{eqnarray}
where
$A$, $B$, $C$, $D$ and $E$ are free parameters fitted to the data with the values reported in Ref.~\cite{hQM-strange}. 

\section{$^3P_0$ pair-creation model}
\label{3P0 model}
\begin{figure}
\centering
   \includegraphics[scale=0.5]{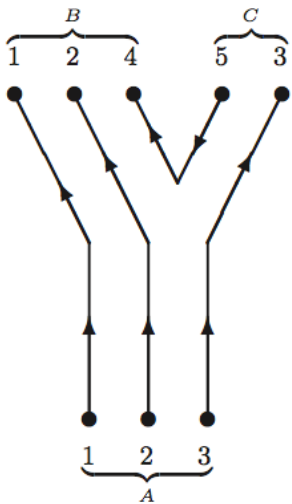}
\caption{The $^3P_0$ pair-creation model of hadron vertices; the $q \bar q$ pair (45) is created in a $^3P_0$ flavor-color singlet. A is the initial-state baryon; B and C are the final baryon and meson states, respectively.}
\label{fig:3P0decay}
\end{figure}

The $^3P_0$ pair-creation model is an effective model to compute open-flavor strong decays \cite{Micu,LeYaouanc,Roberts:1992}. 
Here, a hadron decay takes place in its rest frame and proceeds via the creation of an additional $q \bar q$ pair. 
The quark-antiquark pair is created with the quantum numbers of the vacuum, i.e. $J^{PC} = 0^{++}$ (see Fig. \ref{fig:3P0decay}), and the decay amplitude can be expressed as \cite{Micu,LeYaouanc,Ackleh:1996yt,Barnes:2005pb,Ferretti:2013faa,Ferretti:2013vua,bottomonium}
\begin{equation}
	\label{eqn:3P0-decays-ABC}
	\Gamma_{A \rightarrow BC} = \Phi_{A \rightarrow BC}(q_0) \sum_{\ell} 
	\left| \left\langle BC q_0  \, \ell J \right| T^\dag \left| A \right\rangle \right|^2 \mbox{ }.
\end{equation}

In this paper, we focus on the open-flavor strong decays of light baryons (i.e. made up of $u,d,s$ quarks) in the $^3P_0$ model.   
We assume harmonic oscillator wave functions, depending on a single oscillator parameter $\alpha_b$ for the baryons and $\alpha_c$ for the mesons. 
The final state is characterized by the relative orbital angular momentum $\ell$ between $B$ and $C$ and a total angular momentum 
$\vec{J} = \vec{J}_b + \vec{J}_c + \vec{\ell}$.

\begin{table}
\caption{Effective meson and baryon masses, $\tilde M$ [see Eq. (\ref{eqn:eff-PSF})], from Refs. \cite{Capstick:1992th,Kokoski:1985is}.}
\label{tab:eff-masses}  
\centering
\begin{tabular}{cc} 
\hline 
\hline 
\noalign{\smallskip}
State & $\tilde M$ (GeV) \\ 
\noalign{\smallskip}
\hline 
\noalign{\smallskip}
$N$      & 1.10 \\  
$\Delta$ & 1.10 \\  
$\pi$    & 0.72 \\  
$\rho$   & 0.72 \\
$\eta$   & 0.85 \\ 
$\omega$ & 0.85 \\ 
\noalign{\smallskip}
\hline 
\hline
\end{tabular}
\end{table}

\subsection{Phase space factor}
\label{Phase space factor}
The coefficient $\Phi_{A \rightarrow BC}(q_0)$ is the phase space factor for the decay.
We show three possible prescriptions. The first in the non-relativistic expression,
\begin{equation}
	\label{eqn:nonrel-PSF}
	\Phi_{A \rightarrow BC}(q_0) = 2 \pi q_0 \frac{M_b M_c}{M_a}  \mbox{ },
\end{equation}
depending on the relative momentum $q_0$ between $B$ and $C$ and on the masses of the two intermediate-state hadrons, $M_b$ and $M_c$. The second option is the standard relativistic form,
\begin{equation}
	\label{eqn:rel-PSF}
	\Phi_{A \rightarrow BC}(q_0) = 2 \pi q_0 \frac{E_b(q_0) E_c(q_0)}{M_a}  \mbox{ },
\end{equation}
depending on $q_0$ and on the energies of the two intermediate-state hadrons, $E_b = \sqrt{M_b^2 + q_0^2}$ and $E_c = \sqrt{M_c^2 + q_0^2}$.
The third possibility is to use an effective phase space factor \cite{Capstick:1992th,Kokoski:1985is},
\begin{equation}
	\label{eqn:eff-PSF}
	\Phi_{A \rightarrow BC}(q_0) = 2 \pi q_0 \frac{\tilde M_b \tilde M_c}{M_a}  \mbox{ },
\end{equation}
where $\tilde M_b$ and $\tilde M_c$ are the effective baryon and meson masses, respectively, evaluated by means of a spin-independent interaction (see Table \ref{tab:eff-masses}). According to Ref. \cite{Kokoski:1985is}, this is valid in the weak-binding limit, where $\rho$ and $\pi$ are degenerate and $\tilde m_\pi = 5.1 m_\pi$.

In the case of heavy baryons and mesons, whose internal dynamics is almost non-relativistic and the hyperfine interactions are relatively small, the three types of phase space factors provide almost the same results.

\subsection{Transition operator}
\label{Creation vertex}
The transition operator of the $^{3}P_0$ model is given by \cite{Ferretti:2013faa,Ferretti:2013vua,bottomonium}:
\begin{eqnarray}
T^{\dagger} &=& -3 \gamma_0 \, \int d \vec{p}_4 \, d \vec{p}_5 \, 
\delta(\vec{p}_4 + \vec{p}_5) \, C_{45} \, F_{45} \, V(\vec{p}_4 - \vec{p}_5) 
\nonumber\\
&& \hspace{0.5cm}  \left[ \chi_{45} \, \times \, {\cal Y}_{1}(\vec{p}_4 - \vec{p}_5) \right]^{(0)}_0 \, 
b_4^{\dagger}(\vec{p}_4) \, d_5^{\dagger}(\vec{p}_5)    \mbox{ }.
\label{3p0}
\end{eqnarray}
Here, $\gamma_0$ is the pair-creation strength, and $b_4^{\dagger}(\vec{p}_4)$ and $d_5^{\dagger}(\vec{p}_5)$ are the creation operators for a quark and an antiquark with momenta $\vec{p}_4$ and $\vec{p}_5$, respectively.
The $q \bar q$ pair is characterized by a color-singlet wave function $C_{45}$, a flavor-singlet wave function $F_{45}$, a spin-triplet wave function $\chi_{45}$ with spin $S=1$ and a solid spherical harmonic ${\cal Y}_{1}(\vec{p}_4 - \vec{p}_5)$, since the quark and antiquark are in a relative $P$-wave. 
$V(\vec{p}_4 - \vec{p}_5) = e^{-\alpha_{\rm d}^2 (\vec p_4 - \vec p_5)^2/8}$ is the pair-creation vertex or quark form factor. If one does not consider the quark form factor, i.e. $\alpha_d = 0$, the vertex reduces to a constant ($V = 1$), see Appendix~\ref{Pair-creation vertex}.

In our calculations, we introduce a strange quark-pair suppression mechanism in the baryon sector analogously to what was done in charmonium to suppress unphysical heavy quark pair-creation \cite{Kalashnikova:2005ui,Ferretti:2013faa,Ferretti:2013vua,bottomonium}. The mechanism consists in introducing a reduction factor $m_n/m_q$ in the pair-creation operator which suppresses the creation of strange quark pairs ($q=s$) by a factor of $m_n/m_s$ with respect to the nonstrange quark pairs ($q=u$ and $q=d$). 
This particular choice for the pair-creation strength breaks the $SU(3)$ flavor symmetry and suppresses the creation of $s \bar s$ pairs. Its effect cannot be re-absorbed in a redefinition of the model parameters or in a different choice of the $^3P_0$ model vertex factor.

In Appendix~\ref{Ap1} we present the derivation of the transition amplitudes in the $^{3}P_{0}$ pair-creation model using the Jacobi coordinates of 
Eq.~(\ref{jacobi}) and including the effects of a gaussian smearing of the pair-creation vertex. In Appendix~\ref{Ap2} we present a derivation of the 
flavor coupling coefficients including the strangeness suppression mechanism which is valid for both pseudoscalar and vector mesons.  

\subsection{Mixing between $N(1535)S_{11}$ and $N(1650)S_{11}$}
To better reproduce the experimental data, we introduce a mixing between $N(1535)S_{11}$ and $N(1650)S_{11}$ resonances,
\begin{eqnarray}
	| N(1535)S_{11} \rangle &=& | ^28_{1/2} \rangle \cos \theta + | ^48_{1/2} \rangle \sin \theta \mbox{ },
\nonumber\\
	| N(1650)S_{11} \rangle &=& - | ^28_{1/2} \rangle \sin \theta + | ^48_{1/2} \rangle \cos \theta \mbox{ },
\end{eqnarray}
where $\theta = 38^{\circ}$ is the mixing angle.
This was done in Refs. \cite{Bijker:1994yr,Hey:1974nc}, to correct the disagreement between experimental and theoretical results for the helicity amplitudes of the $N(1535)S_{11}$ and $N(1650)S_{11}$ resonances. 
In Ref.~\cite{Garzon:2014ida} this problem was solved with a mixture of pseudoscalar meson-baryon and vector meson-baryon in a coupled-channels scheme. 

\section{Open-flavor strong decays. Results and discussion}
\label{Strong decay widths} 
In this section, we present the results of our calculations of the open-flavor strong decays of non-strange baryons and hyperons into baryon-pseudoscalar and baryon-vector mesons, using the U(7) model (see Sec. \ref{Open-flavor strong decays calculated using the U(7) model spectrum}) and hQM model (see Sec. \ref{Open-flavor strong decays calculated using the hQM model spectrum}). The decay amplitudes are computed in the $^3P_0$ model of Refs. \cite{Micu,LeYaouanc} and Sec. \ref{3P0 model}.

\subsection{Open-flavor strong decays calculated by using the U(7) model}
\label{Open-flavor strong decays calculated using the U(7) model spectrum}
Here, we show our results for the open-flavor decays by using the U(7) algebraic model of Sec.~\ref{U(7) algebraic model} and Refs. \cite{Bijker:1994yr-first,Bijker:1994yr}.

\begin{table} 
\caption{$^3P_0$ model parameter values used in the calculations, in combination with the relativistic phase space factor of Eq.~(\ref{eqn:rel-PSF}) (column 2) and the effective phase space factor of Eq.~(\ref{eqn:eff-PSF}) (column 3). The parameter values are obtained in a fit to the experimental strong decay widths, see App.~\ref{Pair-creation vertex} 
and Table~\ref{tab:9transitions}, last column. The values of the constituent quark masses $m_n$ ($n = u,d$) and $m_s$ are used in the pair-creation operator of Eq.~(\ref{new3p0}). $\alpha_{b}$ is the harmonic oscillator parameter of baryons $A$ and $B$, $\alpha_{c}$ that of meson $C$ and $\alpha_{d}$ is the quark form factor parameter.}
\label{tab:parameters}  
\centering
\begin{tabular}{cclcl} 
\hline 
\hline 
\noalign{\smallskip}
Parameter           & \multicolumn{2}{c}{Rel. PSF} & \multicolumn{2}{c}{Eff. PSF} \\ 
\noalign{\smallskip}
\hline 
\noalign{\smallskip}
$\gamma_0$   & 14.3 &             & 13.2 & \\  
$\alpha_{b}$ & 2.99 & GeV$^{-1}$  & 2.69 & GeV$^{-1}$ \\  
$\alpha_{c}$ & 2.38 & GeV$^{-1}$  & 2.02 & GeV$^{-1}$ \\
$\alpha_{d}$ & 0.52 & GeV$^{-1}$  & 0.82 & GeV$^{-1}$ \\
$m_n$        & 0.33 & GeV         &      & \\
$m_s$        & 0.55 & GeV         &      & \\
\noalign{\smallskip}
\hline 
\hline
\end{tabular}
\end{table}

\begin{table*}
\caption{Strong decay widths of three- and four-star nucleon resonances (in MeV), calculated with the $U(7)$ algebraic model 
and the hypercentral QM. For the $U(7)$ model the calculation is done for the relativistic and the effective phase space factors 
of Eqs.~(\ref{eqn:rel-PSF}) and (\ref{eqn:eff-PSF}), respectively, in combination with the parameters of Table~\ref{tab:parameters} 
(RPSF and EPSF). For the hypercentral QM we present the results for the relativistic phase space factor (RPSF) in combination 
with the parameters of Table \ref{tab:parameters-3P0+hQM}. The experimental values are taken from Ref.~\cite{Nakamura:2010zzi}. 
Decay channels labeled by -- are below threshold. The symbols ($S$) and ($D$) stand for $S$- and $D$-wave decays, respectively.}
\label{tab:nuc} 
\centering
\begin{tabular}{ccccccccccccc}
\hline
\hline
\noalign{\smallskip}
Model&Resonance & Status & $M$ [MeV] & $N \pi$ & $N \eta$ & $\Sigma K$ & $\Lambda K$ & $\Delta \pi$ & $N \rho$ & $N \omega$ &  \\
\noalign{\smallskip}
\hline
\noalign{\smallskip}                                                        
&$N(1440)P_{11}$          & **** & 1430-1470 & $110-338$ & $0-5$  &      &      &  $22-101$ & & & Exp. \\
U(7)&$^28_{1/2}[56,0^+_2]$ &        & 1444          & 85        & --     & --   & --   & 13  & -- & -- &  RPSF \\
U(7)&$^28_{1/2}[56,0^+_2]$ &        & 1444          & 108      & --     &      &       & 22  &    &     &  EPSF \\ 
hQM&$^{2}8_{ 1/2}[56,0_2^+]$ &       & 1550          & 105       & --     & --   & --   & 12  & -- & -- & RPSF \\  \\                         
&$N(1520)D_{13}$ & **** & 1515-1530 & 102       & 0      &      &       & 342 &    &    & Exp. \\
U(7)&$^28_{3/2}[70,1^-_1]$ &        & 1563 & 134       & 0      & --   & --   & 207 & -- & -- &  RPSF \\
U(7)&$^28_{3/2}[70,1^-_1]$ &        & 1563          & 102       & 0      &      &       & 342 &    &    &  EPSF \\ 
hQM&$^{2}8_{ 3/2}[70,1_1^-]$ &       & 1525          & 111        & 0     & --   & --   & 206 & -- & -- & RPSF \\  \\
&$N(1535)S_{11}$ &**** & 1520-1555 & $44-96$ & $40-91$ & & & $< 2$ & & & Exp. \\
U(7)&$^28_{1/2}[70,1^-_1]$ &        & 1563 &  63 & 75 & -- & --  & 16 & -- & -- &  RPSF \\
U(7)&$^28_{1/2}[70,1^-_1]$ &        & 1563 & 106 & 86 &    &     & 14 &     &     &  EPSF \\   
hQM&$^{2}8_{ 1/2}[70,1_1^-]$ &       & 1525          & 84       & 50    & --   & --   & 6  & -- & -- & RPSF \\ \\
&$N(1650)S_{11}$ & **** & 1640-1680 & $60-162$ & $6-27$ & & $4-20$ & $0-45$ & & & Exp. \\
U(7)&$^48_{1/2}[70,1^-_1]$ &        & 1683 & 41 & 72 & -- & 0 & 18 & -- & -- &  RPSF \\
U(7)&$^48_{1/2}[70,1^-_1]$ &        & 1683 & 71 & 83 & & & 15 & & &  EPSF \\ 
hQM&$^{2}8_{ 1/2}[70,1_2^-]$ &       & 1574          & 51        & 29     & --   & 0  & 4  & -- & -- & RPSF \\   \\
&$N(1675)D_{15}$ & **** & 1670-1685 & $46-74$ & $0-2$ & & $<2$ & $65-99$ & & & Exp. \\
U(7)&$^48_{5/2}[70,1^-_1]$ &        & 1683 & 47 & 11  & -- & 0 & 108 & -- & -- &  RPSF \\
U(7)&$^48_{5/2}[70,1^-_1]$ &        & 1683 & 29 & 7    &    &    & 79   &    &     &  EPSF \\
hQM&$^{4}8_{ 5/2}[70,1_1^-]$ &                   &  1579          & 41        & 9     & --   & --   & 85 & -- & -- & RPSF \\    \\
&$N(1680)F_{15}$ & **** & 1675-1690 & $78-98$ & $0-1$ & & & $6-21$ & & & Exp. \\
U(7)&$^28_{5/2}[56,2^+_1]$ &        & 1737 &   121 & 1 & -- & 0 & 100 & -- & -- &  RPSF \\
U(7)&$^28_{5/2}[56,2^+_1]$ &        & 1737 & 63 & 0    &    &    & 99   &    &     &  EPSF \\
hQM&$^{2}8_{ 5/2}[56,2_1^+]$ &                   &  1798          & 91       & 0     & 0   & 0   & 92 & -- & -- & RPSF \\    \\
&$N(1700)D_{13}$ &  *** & 1650-1750 & $7-43$ & $0-3$ & & $<8$ & $10-225$ ($S$) & & & Exp. \\
  &                           &       &                    &              &              &           &       & $<50$ ($D$) &  \\
U(7)&$^48_{3/2}[70,1^-_1]$ &        & 1683 & 9 & 3 & -- & 0 & 561 & -- & -- &  RPSF \\
U(7)&$^48_{3/2}[70,1^-_1]$ &        & 1683 & 5 & 2    &    &    & 657  &    &     &  EPSF \\
hQM&$^{2}8_{ 3/2}[70,1_2^-]$ &                   &  1606    & 0      & 0    & 0  & 0   & 0  & -- & -- & RPSF \\    \\     
&$N(1710)P_{11}$ &  *** & 1680-1740 & $3-50$ & $5-75$ & & $3-63$ & $8-100$ & $3-63$ & & Exp. \\ 
U(7)&$^28_{1/2}[70,0^+_1]$ &        & 1683 & 5 &  9 & 0 & 3 & 56 & -- & -- &  RPSF \\
U(7)&$^28_{1/2}[70,0^+_1]$ &        & 1683 & 11 & 9    &    &    & 58  &    &     &  EPSF \\
hQM&$^{2}8_{ 1/2}[70,0_1^+]$ &                   &  1808   & 18       & 12   & 0   & 14.1   & 70  & -- & -- & RPSF \\   \\  
&$N(1720)P_{13}$ & **** & 1650-1750 & $12-56$ & $5-20$ & & $2-60$ & $90-360$ & $105-340$ & & Exp. \\ 
U(7)&$^28_{3/2}[56,2^+_1]$ &        & 1737 &  111 &  7   &  0 & 14 & 36 & 5 & 0 &  RPSF \\
U(7)&$^28_{3/2}[56,2^+_1]$ &        & 1737 & 123 & 7    &    &    & 28  &    &     &  EPSF  \\  
hQM&$^{2}8_{ 3/2}[56,2_1^+]$ &                   &  1797   & 141      & 8     & 0   & 12  & 30 & 77 & 5 & RPSF \\ \\                    
&$N(1875)D_{13}$ &*** & 1820-1920 & $3-70$ & $0-22$ & $0-4$ &       & $48-192$ ($S$) & $0-38$ & $22-90$ & Exp. \\
   &                          &     &                    &             &              &           &       & $11-86$ ($D$)   &             &              & Exp. \\
U(7)&$^48_{3/2}[70,2^-_1]$ &        & 1975 & 0 & 0 & 0 & 0 & 0 & 0 & 0 &  RPSF \\
U(7)&$^48_{3/2}[70,2^-_1]$ &        & 1975 & 0 & 0    &    &    & 0  &    &     &  EPSF \\ 
hQM&$^{4}8_{ 3/2}[70,1_1^-]$ &                   &  1899   & 14       & 8    & 2   & 0   &560 & 80 & 82 & RPSF \\   \\
&$N(1900)P_{13}$ &*** & 1875-1935 & $20-37$ & $24-44$ & $6-26$ & $0-37$ &  &  & $75-120$ & Exp. \\
U(7)&$^28_{3/2}[70,2^+_1]$ &        & 1874 & 11 & 12 & 1 & 13 & 63 & 64 & 24 &  RPSF \\
U(7)&$^28_{3/2}[70,2^+_1]$ &        & 1874 & 17 & 11    &    &    & 33  &    &     &  EPSF \\ 
hQM&$^{2}8_{ 3/2}[70,2_1^+]$ &                   &  1853   & 15       & 12    & 1   & 13& 70 & 53 & 23 & RPSF \\ 
\noalign{\smallskip}
\hline
\hline
\end{tabular}
\end{table*}

\begin{table*}
\caption{As Table~\protect\ref{tab:nuc}, but for $\Delta$ resonances. The symbols ($S$), ($P$), ($D$) and ($F$) stand for $S$-, $P$-, $D$- and $F$-wave decays, respectively.}
\label{tab:del}
\centering 
\begin{tabular}{ccccccccccc}
\hline
\hline
\noalign{\smallskip}
Model&Resonance & Status & $M$ [MeV] & $N \pi$ & $\Sigma K$ & $\Delta \pi$ & $\Delta \eta$ & $\Sigma^* K$ & $N \rho$ & \\
\noalign{\smallskip}
\hline
\noalign{\smallskip}
&$\Delta(1232)P_{33}$ & **** & 1230-1234 & $114-120$ & & & & & & Exp. \\
U(7)&$^410_{3/2}[56,0_1^+]$ &    & 1246          & 71 & -- & -- & -- & -- & -- & RPSF \\
U(7)&$^410_{3/2}[56,0_1^+]$ &    & 1246          & 115 &   & -- & -- &  &  & EPSF \\
hQM&$^410_{3/2}[56,0_1^+]$ &    & 1240          & 63& -- & -- & -- & -- & -- & RPSF \\   \\
&$\Delta(1600)P_{33}$ &  *** & 1550-1700 & $22-105$ & & $88-294$ & & & $< 88$ & Exp. \\
U(7)&$^410_{3/2}[56,0_2^+]$ &    & 1660          &  17   & -- & 65 & -- & -- & -- & RPSF \\
U(7)&$^410_{3/2}[56,0_2^+]$ &    & 1660          & 24 &   & 74 & -- &  &  & EPSF \\ 
hQM&$^{4}10_{3/2}[56,0_2^+]$ &    & 1727          & 31 & -- & 69 & -- & -- & -- & RPSF \\ \\
&$\Delta(1620)S_{31}$ & **** & 1615-1675 & $26-45$ & & $39-90$ & &  & $9-38$ & Exp. \\
U(7)&$^210_{1/2}[70,1_1^-]$ &    & 1649          & 5 & -- & 76 & -- & -- & -- & RPSF \\
U(7)&$^210_{1/2}[70,1_1^-]$ &    & 1649          & 10 &   & 61 & -- &  &  & EPSF \\ 
hQM&$^{2}10_{1/2}[70,1_1^-]$ &    & 1584          & 9 & -- & 59& -- & -- & -- & RPSF \\  \\
&$\Delta(1700)D_{33}$ & **** & 1670-1770 & $20-80$ & & $50-200$ ($S$) & $6-28$ & &$48-165$ & Exp. \\
 &                                  &        &                    &                & & $10-60$ ($D$) & & &  \\
U(7)&$^210_{3/2}[70,1_1^-]$ &    & 1649          & 46 & -- & 311 & -- & -- & -- & RPSF \\
U(7)&$^210_{3/2}[70,1_1^-]$ &    & 1649          & 27 &   & 343 & -- &  &  & EPSF \\
hQM&$^{2}10_{3/2}[70,1_1^-]$ &    & 1584          & 40& -- & 333 & -- & -- & -- & RPSF \\   \\ 
&$\Delta(1905)F_{35}$ & **** & 1855-1910 & $24-60$ & & $< 100$ & & & $> 162$ & Exp. \\
U(7)&$^410_{5/2}[56,2_1^+]$ &    & 1921          & 31 & 1 & 188 & 19 & 0 & 99 & RPSF \\
U(7)&$^410_{5/2}[56,2_1^+]$ &    & 1921          & 14 &   & 139 & 14 &  &  & EPSF \\
hQM&$^{4}10_{5/2}[56,2_1^+]$ &    & 1844          & 26& 0 &182 &15 & -- & 88 & RPSF \\    \\                                  
&$\Delta(1910)P_{31}$ & **** & 1860-1910 & $33-102$ & $9-48$ & $70-299$ & & & & Exp. \\
U(7)&$^410_{1/2}[56,2_1^+]$ &    & 1921          & 26 & 38 & 32 & 4 & -- & 64 & RPSF \\
U(7)&$^410_{1/2}[56,2_1^+]$ &    & 1921          & 39 &   & 27 & 3 &  &  & EPSF \\ 
hQM&$^410_{1/2}[56,2_1^+]$ &    & 1871          & 49 & 38 &34 & 4 & -- & 60 & RPSF \\ \\
&$\Delta(1920)P_{33}$ &  *** & 1900-1970 & $9-60$ & $3-7$ & $18-102$ ($P$) & $13-69$ & & 0 & Exp. \\
    &                                &       &                   &              &           & $45-195$ ($F$)  & & \\
U(7)&$^410_{3/2}[56,2_1^+]$ &    & 1921          & 7 & 23 & 132 & 22 & 5 & 105 & RPSF \\
U(7)&$^410_{3/2}[56,2_1^+]$ &    & 1921          & 14 &   & 96 & 15 &  &  & EPSF \\
hQM&$^{4}10_{3/2}[56,2_1^+]$ &    & 1856        & 17& 22 & 137 & 20 & -- & 102 & RPSF \\  \\
&$\Delta(1930)D_{35}$ &  *** & 1920-1970 & $11-75$ & & & & & & Exp. \\
U(7)&$^210_{5/2}[70,2_1^-]$ &    & 1946          & 0 & 0 & 0 & 0 & 0 & 0 & RPSF \\
U(7)&$^210_{5/2}[70,2_1^-]$ &    & 1946          & 0 &   & 0 & 0 &  &  & EPSF \\  \\
&$\Delta(1950)F_{37}$ & **** & 1940-1960 & $82-151$ & $1-2$ & $47-101$ & &  & $< 34$ & Exp. \\
U(7)&$^410_{7/2}[56,2_1^+]$ &    & 1921          & 172 & 5 & 92 & 1 & 0 & 22 & RPSF \\
U(7)&$^410_{7/2}[56,2_1^+]$ &    & 1921          & 72 &   & 40 & 1 &  &  & EPSF \\
hQM&$^{4}10_{7/2}[56,2_1^+]$ &    & 1851        &146 & 3 & 70 & 1 & -- & 16 & RPSF \\
\noalign{\smallskip}
\hline
\hline
\end{tabular}
\end{table*}

\begin{table*}
\caption{As Table \ref{tab:nuc}, but for $\Sigma$ and $\Sigma^*$ resonances.} 
\label{tab:sig}
\centering
\begin{tabular}{ccccccccccccccccc}
\hline
\hline
\noalign{\smallskip}
Model&Baryon & Status & $M$ [MeV] & $N \overline{K}$ & $\Sigma \pi$ & $\Lambda \pi$ & $\Sigma \eta$ & $\Xi K$ & $\Delta \overline{K}$ & $\Sigma^* \pi$ & $N \overline{K}^*$ & $\Sigma \rho$ & $\Lambda \rho$ & $\Sigma \omega$ & \\
\noalign{\smallskip}
\hline
\noalign{\smallskip}
&$\Sigma(1660)P_{11}$ &  *** & 1630-1690 & $4-60$ & seen & seen & & & & &&&&& Exp. \\
{$ $ $U(7)$}&$^28_{1/2}[56,0^+_2]$ &        & 1604 & 3 & 38 & 14 & -- & -- & -- & 7 & -- & -- & -- & -- & RPSF \\ 
\hline
{$ $ hQM}&{$ $$^28_{1/2}[56,0^+_2]$} &        & {$ $1704} & {$ $ 4} & {$ $ 45}  & {$ $ 18} & -- & -- & -- & {$ $ 6}& -- & -- & -- & -- & RPSF  \\
&$\Sigma(1670)D_{13}$ & **** & 1665-1685 & $3-10$ & $12-48$ & $2-12$ & & & & &&&&& Exp. \\
{$ $ $U(7)$}&$^48_{3/2}[70,1^-_1]$ &         & 1711 & 5 & 78 & 8 & -- & -- & -- & 36 & -- & -- & -- & -- & RPSF \\ 
{$ $ hQM}&{$ $$^28_{3/2}[70,1^-_1]$}  &        & {$ $1799 }& {$ $ 4} & {$ $ 62}  & {$ $ 7} & -- & -- & -- & {$ $ 29}& -- & -- & -- & -- & RPSF  \\
&$\Sigma(1750)S_{11}$ &  *** & 1730-1800 & $6-64$ & $< 13$ & seen & $9-88$ & & & &&&&& Exp. \\
{$ $ $U(7)$}&$^48_{1/2}[70,1^-_1]$ &         & 1711  & 3 & 109 & 3 & 28& -- & -- & 9 & -- & -- & -- & -- &  RPSF \\
{$ $ hQM}&{$ $$^28_{1/2}[70,1^-_1]$ } &        & {$ $1799} & {$ $ 5} & {$ $ 151}  & {$ $ 6} & {$ $ 27}  & -- & -- & {$ $ 7}& -- & -- & -- & -- & RPSF \\   
&$\Sigma(1775)D_{15}$ & **** & 1770-1780 & $39-58$ & $2-7$ & $15-27$ & & & & $8-16$ & & & && Exp. \\                                   
{$ $ $U(7)$}&$^48_{5/2}[70,1^-_1]$ &         & 1822 & 101 & 17 & 38 & { 0} & -- & 4 & 11 & -- & -- & -- & -- &  RPSF \\ 
{$ $ hQM}&{$ $$^48_{5/2}[70,1^-_1]$}  &        & {$ $1914} & {$ $ 79} & {$ $ 13}  & {$ $ 30} &{$ $ 0}   &  -- &  {$ $ 2}  & {$ $ 7}& -- & -- & -- & -- & RPSF  \\                                                                  
&$\Sigma(1915)F_{15}$ & **** & 1900-1935 & $4-24$ & seen & seen & & & & $< 8$ & & & && Exp. \\ 
{$ $ $U(7)$}&$^28_{5/2}[56,2^+_1]$ &         & 1872 &  6 & 58 & 33 & 2 & 1 & 96 & 23 & 6 & -- & 2 & -- &  RPSF \\  
{$ $ hQM}&{$ $$^28_{5/2}[56,2^+_1]$ }&        & {$ $1906 }& {$ $ 4} & {$ $ 44}  & {$ $ 26} & {$ $ 1}  & {$ $ 0} & {$ $ 88} & {$ $ 22}&  {$ $ 5} & -- &  {$ $ 2}  & -- & RPSF  \\                                                                   
&$\Sigma(1940)D_{13}$   &  *** & 1900-1950 & $< 60$ & seen & seen & & & seen & seen & seen & & & &  Exp. \\ 
{$ $ $U(7)$}&$^28_{3/2}[56,1^-_1]$ &         & 1974 & 0 & 0 & 0 & { 0} & 0 & 0 & 0 & 0 & -- & 0 & -- &  RPSF \\
 {$ $ hQM}&{$ $ $^48_{3/2}[70,1^-_1]$ } &        & {$ $1914} & {$ $ 31} & {$ $ 6}  & {$ $ 11} & {$ $ 0}  &  {$ $ 0} &  {$ $ 554} & {$ $ 99}&  {$ $ 251}  & -- &  {$ $ 0} & -- & RPSF   \\                                                                
&$\Sigma^*(1385)P_{13}$ & **** & 1383-1385 &   & $30-32$ & $4-5$ &   &  &   &  &   &   &  &   & Exp. \\ 
{$ $ $U(7)$}&$^410_{3/2}[56,0^+_1]$ &         & 1382 & -- & 3 & 27 & -- & -- & -- & -- & -- & -- & -- & --  & RPSF \\ 
{$ $ hQM}&{$ $$^410_{3/2}[56,0^+_1]$} &        & {$ $1372} & --& {$ $ 3}  & {$ $ 24} & --  & -- & -- &--& -- & -- & -- & -- & RPSF  \\
&$\Sigma^*(2030)F_{17}$   & **** & 2025-2040 & $26-46$ & $8-20$ & $26-46$ & & $<4$ & $15-40$ & $8-30$ &  &  &  & & Exp. \\ 
{$ $ $U(7)$}&$^410_{7/2}[56,2^+_1]$ &         & 2012 & 54 & 37 & 75 & {$ $ 8} & 1 & 30 & 37 & 7 & 0 & 4 & 0 & RPSF \\   
{$ $ hQM}&{$ $$^410_{7/2}[56,2^+_1]$ } &        & {$ $2085} & {$ $ 44} & {$ $ 30}  & {$ $ 62} & {$ $ 6}  & {$ $ 1} & {$ $ 22} & {$ $ 27}& {$ $ 5} & {$ $ 0} & {$ $ 2}& {$ $ 0} & RPSF\\
\noalign{\smallskip}
\hline
\hline
\end{tabular}
\end{table*}

\begin{table*}
\caption{As Table \ref{tab:nuc}, but for $\Lambda$ and $\Lambda^*$ resonances. }
\label{tab:lam}
\centering 
\begin{tabular}{ccccccccccccc}
\hline
\hline
\noalign{\smallskip}
Model&Baryon & Status & $M$ [MeV] & $N \overline{K}$ & $\Sigma \pi$ & $\Lambda \eta$ & $\Xi K$ & $\Sigma^* \pi$ & $N \overline{K}^*$ & $\Sigma \rho$ & $\Lambda \omega$ & \\
\noalign{\smallskip}
\hline
\noalign{\smallskip}
&$\Lambda(1600)P_{01}$   &  *** & 1560-1700 & $8-75$ & $5-150$ &    &     &      &    &     &  & Exp. \\
{$ $ $U(7)$}&$^{2}8_{ 1/2}[56,0_2^+]$       &       & 1577          & 79 & 40 & -- & -- & 19 & -- & -- & -- & RPSF \\
{$ $ hQM}& {$ $$^{2}8_{ 1/2}[56,0_2^+]$ } &        & {$ $1627} & {$ $93} & {$ $ 46}  & {$ $ --} & {$ $ --}  & {$ $ 18} & {$ $ --} & {$ $ --}& {$ $ --}& {$ $RPSF}  \\
&$\Lambda(1670)S_{01}$   & **** & 1660-1680 & $5-15$ & $6-28$ & $3-13$ & &    &   &  &  & Exp. \\
{$ $ $U(7)$}&$^{2}8_{ 1/2}[70,1_1^-]$       &       & 1686          & 217 & 33 & {9} & -- & 7 & -- & -- & -- & RPSF \\ 
{$ $ hQM}& {$ $$^{2}8_{ 1/2}[70,1_1^-]$ } &        & {$ $1722} & {$ $ 267} & {$ $ 39}  & {$ $ 0} & {$ $ --}  & {$ $ 5} & {$ $ --} & {$ $ --}& {$ $ --}& {$ $RPSF}   \\
&$\Lambda(1690)D_{03}$   & **** & 1685-1690 & $10-21$ & $10-28$ & & &    &   &  &  & Exp. \\
{$ $ $U(7)$}&$^{2}8_{ 3/2}[70,1_1^-]$       &       & 1686          & 150 & 16 & { 0} & -- & 168 & -- & -- & -- &  RPSF \\ 
{$ $ hQM}&{$ $$^{2}8_{ 3/2}[70,1_1^-]$  } &        & {$ $1722} & {$ $ 119} & {$ $ 23}  & {$ $ 0} & {$ $ --}  & {$ $ 168} & {$ $ --} & {$ $ --}& {$ $ --}& {$ $RPSF}   \\
&$\Lambda(1800)S_{01}$   &  *** & 1720-1850 & $50-160$ & seen & & & seen    &   &  &  & Exp. \\
{$ $ $U(7)$}&$^{4}8_{ 1/2}[70,1_1^-]$       &       & 1799          & 0 & 67 & { 85} & -- & 13 & -- & -- & -- &  RPSF \\ 
{$ $ hQM}& {$ $$^{4}8_{ 1/2}[70,1_1^-]$  } &        & {$ $1837} & {$ $0} & {$ $ 98}  & {$ $ 90} & {$ $ --}  & {$ $ 10} & {$ $ --} & {$ $ --}& {$ $ --}& {$ $RPSF}  \\
&$\Lambda(1810)P_{01}$   &  *** & 1750-1850 & $10-125$ & $5-100$ & & & seen    &   &  &  & Exp. \\
 {$ $ $U(7)$}&    $^{2}8_{ 1/2}[70,0_1^+]$  &       & 1799          & 16 & 4 & { 2} & -- & 40 & -- & -- & --  &  RPSF \\ 
 {$ $ hQM}& {$ $$^{2}8_{ 1/2}[56,0_3^+]$ } &        & {$ $1973} & {$ $ 0} & {$ $ 0}  & {$ $ 0} & {$ $ --}  & {$ $ 0} & {$ $ --} & {$ $--}& {$ $ --}& {$ $RPSF}  \\
&$\Lambda(1820)F_{05}$   & **** & 1815-1825 & $39-59$ & $6-13$ & & & $4-9$    &   &  &  & Exp. \\
{$ $ $U(7)$}&$^{2}8_{ 5/2}[56,2_1^+]$       &       & 1849          & 78 & 31 & {1} & 0 & 73 & -- & -- & -- &  RPSF \\
{$ $ hQM}& {$ $$^{2}8_{ 5/2}[56,2_1^+]$  } &        & {$ $1829} & {$ $ 57} & {$ $ 22}  & {$ $ 0} & {$ $ 0}  & {$ $ 66} & {$ $ --} & {$ $ --}& {$ $ --}& {$ $RPSF}   \\
&$\Lambda(1830)D_{05}$   & **** & 1810-1830 & $2-11$ & $21-83$ & & & $>9$    &   &  &  & Exp. \\
{$ $ $U(7)$}&$^{4}8_{ 5/2}[70,1_1^-]$       &       & 1799          & 0 & 99 & { 9} & 0 & 82 & -- & -- & -- &   RPSF \\  
{$ $ hQM}&{$ $$^{4}8_{ 5/2}[70,1_1^-]$ } &        & {$ $1837} & {$ $ 0} & {$ $ 84}  & {$ $ 7} & {$ $0}  & {$ $ 64} & {$ $ --} & {$ $ --}& {$ $ --}& {$ $RPSF}  \\
&$\Lambda(1890)P_{03}$   & **** & 1850-1910 & $12-70$ & $2-20$ & & & seen  &  &  &  & Exp. \\
{$ $ $U(7)$}&$^{2}8_{ 3/2}[56,2_1^+]$       &       & 1849          & 96 & 69 & { 31} & 2 & 30 & 28 & -- & -- &  RPSF \\
{$ $ hQM}&  {$ $$^{2}8_{ 3/2}[56,2_1^+]$   } &        & {$ $1829} & {$ $ 120} & {$ $ 79}  & {$ $ 11} & {$ $ 1}  & {$ $ 24} & {$ $ 27} & {$ $ --}& {$ $ --}& {$ $RPSF}  \\
&$\Lambda(2110)F_{05}$   & **** & 2090-2140 & $8-63$ & $15-100$ & & & seen  &  &  &  & Exp. \\
{$ $ $U(7)$}&$^{4}8_{ 5/2}[70,2_1^+]$       &       & 2074          & 0 & 14 & {3} & 1 & 136 & 0 & 38 & { 16} &  RPSF \\
{$ $ hQM}& {$ $$^{2}8_{ 5/2}[70,2_1^+]$  } &        & {$ $1995} & {$ $ 167} & {$ $ 20}  & {$ $ 0} & {$ $ 3}  & {$ $ 69} & {$ $ 15} & {$ $ 79}& {$ $ 25}& {$ $RPSF}   \\
&$\Lambda^*(1405)S_{01}$   & **** & 1402-1410 & & $48-52$ & & &  &  &  &  & Exp. \\
{$ $ $U(7)$}&$^{2}1_{ 1/2}[70,1_1^-]$       &       & 1641          & -- & 230 & -- & -- & -- & -- & -- & -- &  RPSF \\  
{$ $ hQM}&{$ $$^{2}1_{ 1/2}[70,1_1^-]$   } &        & {$ $1658} & {$ $ --} & {$ $ 222}  & {$ $ --} & {$ $ --}  & {$ $--} & {$ $ --} & {$ $ --}& {$ $ --}& {$ $RPSF}  \\
&$\Lambda^*(1520)D_{03}$   & **** & 1518-1520 & $6-8$ & $6-7$ & & & 1   &  &  &  & Exp. \\
{$ $ $U(7)$}&$^{2}1_{ 3/2}[70,1_1^-]$       &       & 1641          & 10 & 17 & -- & -- & -- & -- & -- & -- &  RPSF \\   
{$ $ hQM}&{$ $$^{2}1_{ 3/2}[70,1_1^-]$ } &        & {$ $1658} & {$ $ 8} & {$ $ 13}  & {$ $ --} & {$ $ --}  & {$ $--} & {$ $ --} & {$ $ --}& {$ $ --}& {$ $RPSF}  \\
\noalign{\smallskip}
\hline
\hline
\end{tabular}
\end{table*}

\begin{table*}
\caption{As Table \ref{tab:nuc}, but for $\Xi$ and $\Xi^*$resonances.}
\label{tab:xi} 
\begin{tabular}{ccccccccc}
\hline
\hline
\noalign{\smallskip}
Model&Baryon & Status & $M$ [MeV] & $\Sigma \overline{K}$ & $\Lambda \overline{K}$ & $\Xi \pi$ & $\Xi^* \pi$ &  \\
\noalign{\smallskip}
\hline
\noalign{\smallskip}
&$\Xi(1690)S_{11}$       & *** & 1680-1700 &   &   &  & & Exp.  \\
{$ $ $U(7)$}&$^28_{1/2}[70,1^-_1]$ &      & 1828           & 58 & 85 & 14 & 0 &  RPSF  \\ 
{$ $ hQM}&{$ $$^{2}8_{ 1/2}[70,1_1^-]$ } &        & {$ $1938} & {$ $ 55} & {$ $86}  & {$ $ 15} & {$ $ 0}  & {$ $RPSF}  \\
&$\Xi(1820)D_{13}$    &  *** & 1818-1828 & $2-18$ & $3-12$ & $0-8$  & $2-18$ & Exp.  \\
{$ $ $U(7)$}&$^28_{3/2}[70,1^-_1]$ &      & 1828           & 38 & 26 & 6 & 55 &  RPSF  \\ 
{$ $ hQM}&{$ $$^{2}8_{ 3/2}[70,1_1^-]$ } &        & {$ $1938} & {$ $ 29} & {$ $ 21}  & {$ $ 5} & {$ $ 55}  & {$ $RPSF}  \\
&$\Xi^*(1530)P_{13}$         & **** & 1531-1532 & & & $9-10$ & & Exp.  \\
{$ $ $U(7)$}&$^{4}10_{3/2}[56,0_1^+]$ &      & 1524           & -- & -- & 11 & -- &  RPSF  \\
{$ $ hQM}&{$ $$^{4}10_{3/2}[56,0_1^+]$  } &        & {$ $1511} & {$ $ --} & {$ $ --}  & {$ $ 9} & {$ $ --}  & {$ $RPSF} 
\\  
\noalign{\smallskip}
\hline
\hline
\end{tabular}
\end{table*}

The strong decay widths are computed in the $^{3}P_{0}$ model using Eqs.~(\ref{eqn:3P0-decays-ABC}), (\ref{eqn:Mabc}) and (\ref{eqn:epsilon}), by considering two possible choices for the phase space factor: the standard relativistic form of Eq. (\ref{eqn:rel-PSF}) and the effective phase space factor of Eq. (\ref{eqn:eff-PSF}).
The results obtained with the relativistic phase space factor (RPSF) and the model parameters of the second column of Table \ref{tab:parameters} are reported in Tables \ref{tab:nuc}--\ref{tab:omega}; those obtained with the effective phase space factor (EPSF) and the model parameters of the third column of Table \ref{tab:parameters} are reported in Tables \ref{tab:nuc}--\ref{tab:del}. 
The $^3P_0$ model parameters of Table \ref{tab:parameters} are fitted to a sample of 9 transitions, as discussed in App. \ref{Pair-creation vertex}.
In our calculations, whenever available we use the experimental values for the masses of the decaying resonances from the PDG \cite{Nakamura:2010zzi}, otherwise the theoretical predictions of Sec.~\ref{U(7) algebraic model}.  

\begin{table*}
\caption{Strong decay widths of missing nucleon resonances (in MeV) calculated in the U(7) Model of Sec.~\ref{U(7) algebraic model} 
and Refs.~\cite{Bijker:1994yr-first,Bijker:1994yr} (top) and the hypercentral QM of Sec.~\ref{hQM} and Refs.~\cite{pl,chin} (bottom). 
The calculations are carried out using the model parameters of Table~\ref{tab:parameters} (second column) and Table~\ref{tab:parameters-3P0+hQM}, 
respectively. in combination with the relativistic phase space factor of Eq.~(\ref{eqn:rel-PSF}).
Tentative assignments of one and two star resonances are labeled by $^{\ddagger}$.} 
\label{tab:missing-nuc}
\centering 
\begin{tabular}{ccccccccccccc}
\hline
\hline
\noalign{\smallskip}
$N$ & Mass & $N \pi$ & $N \eta$ & $\Sigma K$ & $\Lambda K$ & $\Delta \pi$ & $\Sigma^* K$ 
& $N \rho$ & $N \omega$ & $\Sigma K^*$ & $\Lambda K^*$ & $\Delta \rho$ \\
\noalign{\smallskip}
\hline
\\
\multicolumn{13}{c}{$ $ $U(7)$ Model}\\
\multicolumn{13}{c}{$ $ \line(1,0){300}}\\
\noalign{\smallskip}
$^{2}8_{J}  [20,1_1^+]$ & 1713 & 0 &  0 &  0 &  0 & 0 & -- & -- & -- & -- & -- & -- \\
$^{4}8_{3/2}[70,0_1^+]$ & 1796 & 0 & 3 & 5 & 0 & 65 & -- & 7 & 7 & -- & -- & -- \\
$^{2}8_{5/2}[70,2_1^+]$ & 1874 $^{\ddagger}$ & 106 & 10 & 0 & 3 & 79 & -- & 161 & 8 & -- & -- & -- \\
$^{2}8_{J}  [70,2_1^-]$ & 1874 &   0 &  0 &  0 &  0 &   0 & -- &   0 &  0 & -- & -- & -- \\
$^{4}8_{1/2}[70,2_1^+]$ & 1975 $^{\ddagger}$ & 1  & 8  & 23 & 0 & 19 & 1 & 9 & 9 & -- & -- & -- \\
$^{4}8_{3/2}[70,2_1^+]$ & 1975 $^{\ddagger}$ & 0  & 4  & 11 & 0 & 109 & 5 & 14 & 14 & -- & -- & -- \\
$^{4}8_{5/2}[70,2_1^+]$ & 1975 $^{\ddagger}$ & 6  & 3  & 1  & 0 & 176 & 6 & 16 & 16 & -- & -- & -- \\
$^{4}8_{7/2}[70,2_1^+]$ & 1975 $^{\ddagger}$ & 25 & 13 & 4  & 0 &  99 & 0 & 5 & 4 & -- & -- & -- \\
$^{4}8_{J}  [70,2_1^-]$ & 1975 $^{\ddagger}$ &  0 &  0 &  0 & 0 &   0 &  0 & 0 & 0 & -- & -- & -- \\
$^{2}8_{1/2}[56,1_1^-]$ & 2094 & 5 & 1 & 1 & 5 & 3 & 2 & 48 & 6 & 2 & 2 & 14 \\
$^{2}8_{3/2}[56,1_1^-]$ & 2094 $^{\ddagger}$ & 27 & 0 & 0 & 1 & 23 & 1 & 53 & 11 & 0 & 2 & 13 \\
$^{2}8_{1/2}[70,1_2^-]$ & 1829 $^{\ddagger}$ & 42 & 7 & 0 & 1 & 38 & --  & 0 & 0 & -- & -- & -- \\
$^{4}8_{1/2}[70,1_2^-]$ & 1933 & 8 & 12 & 3 & 0 & 0 & 0 & 0 & 0 & -- & -- & -- \\
$^{4}8_{3/2}[70,1_2^-]$ & 1933 & 0 & 0 & 3 & 0 & 0 & 0 & 0 & 0 & -- & -- & -- \\
$^{4}8_{5/2}[70,1_2^-]$ & 1933 & 0 & 2 & 5 & 0 & 1 & 0 & 0 & 0 & -- & -- & -- \\
\noalign{\smallskip}
\multicolumn{13}{c}{$ $ \line(1,0){300}}\\
\multicolumn{13}{c}{$ $ hQM}\\
\multicolumn{13}{c}{$ $ \line(1,0){300}}\\
\noalign{\smallskip}
$^{4}8_{3/2}[70,2_1^+]$ & 1835 & 4 & 8 & 7 & 0 & 97 & - & 8 & 7 & - & -& - \\
$^{2}8_{1/2}[20,1_1^+]$ & 1836 & 0 & 0 & 0 & 0 & 0 & - & 0 & 0 & - & - & - \\
$^{2}8_{3/2}[20,1_1^+]$ & 1836 & 0 & 0 & 0 & 0 & 0 & - & 0 & 0 & - & - & - \\
$^{4}8_{1/2}[70,2_1^+]$ & 1839 $^{\ddagger}$ & 8 &16 & 15& 0 & 27& - & 6 & 5 & - & - & - \\
$^{4}8_{7/2}[70,2_1^+]$ & 1840 $^{\ddagger}$ & 12& 4 & 0 & 0 & 25& - & 0 & 0 & - & - & - \\
$^{4}8_{5/2}[70,2_1^+]$ & 1844 $^{\ddagger}$ & 3 & 1 & 0 & 0 &137& - & 9 & 8 & - & - & - \\
$^{4}8_{5/2}[70,2_1^+]$ & 1851 $^{\ddagger}$ & 3 & 1 & 0 & 0 &137& - & 9 & 9 & - & - & - \\
$^{4}8_{3/2}[70,0_1^+]$ & 1863 $^{\ddagger}$ & 0 & 4 & 22 & 0 & 83 & - & 12 & { 12} & - & - & - \\
$^{4}8_{1/2}[70,1_1^-]$ & 1887 $^{\ddagger}$ & 0 & 22&119& 0 &87 & - & 32 & 32 & - & - & - \\
$^{4}8_{1/2}[70,1_2^-]$ & 1937 & 0 & 0 & 0 & 0 & 0 & - & 0 & 0 & - & - & - \\
$^{4}8_{5/2}[70,1_2^-]$ & 1942 $^{\ddagger}$ & 0 &0  & 0 & 0& 0 & 0 & 0 & 0 & - & - & - \\
$^{2}8_{1/2}[56,0_3^+]$ & 1943 $^{\ddagger}$ & 0& 0 & 0 & 0 & 0 & 0 & 0 & 0 & - & - &-  \\
$^{4}8_{3/2}[70,1_2^-]$ & 1969 & 0 & 0 & 0 & 0 & 0 & 0 & 0 & 0 & - & - & - \\
\noalign{\smallskip}
\hline
\hline
\end{tabular}
\end{table*}

\begin{table}
\caption{As Table \ref{tab:missing-nuc}, but for missing $\Delta$ resonances. } 
\label{tab:delta-missing} 
\begin{tabular}{cccccccc}
\hline
\hline
\noalign{\smallskip}
$\Delta$ & Mass & $N \pi$ & $\Sigma K$ & $\Delta \pi$ & $\Delta \eta$ & $\Sigma^* K$ & $N \rho$ \\
\noalign{\smallskip}
\hline
\\
\multicolumn{8}{c}{$ $$U(7)$ Model}\\
\multicolumn{8}{c}{$ $ \line(1,0){200}}\\
\noalign{\smallskip}
$^{2}10_{1/2}[70,0_1^+]$ & 1764 $^{\ddagger}$ & 0 & 1 & 70 & -- & -- & 23 \\
$^{2}10_{3/2}[70,2_1^-]$ & 1946 &   0 &  0 &   0 &  0 &  0 & 0  \\
$^{2}10_{3/2}[70,2_1^+]$ & 1947 &   1 &  3 & 106 &  5 &  1 & 80 \\
$^{2}10_{5/2}[70,2_1^+]$ & 1947 $^{\ddagger}$ &   18 &  1 & 107 &  18 &  4 & 32 \\
$^{2}10_{1/2}[70,1_2^-]$ & 1904 $^{\ddagger}$ & 0 & 0 & 0 & 0 & 0 & 0 \\
$^{2}10_{3/2}[70,1_2^-]$ & 1904 & 0 & 0 & 0 & 0 & 0 & 0 \\
\noalign{\smallskip}
\multicolumn{8}{c}{$ $ \line(1,0){200}}\\
\multicolumn{8}{c}{$ $ hQM}\\
\multicolumn{8}{c}{$ $ \line(1,0){200}}\\
\noalign{\smallskip}
$^{2}10_{1/2}[70,0_1^+]$ & 1832 $^{\ddagger}$ & 0 & 2 & 89 & { 7} & - & 57 \\
$^{2}10_{3/2}[70,2_1^+]$ & 1843 & 4 & 1 & 43& 1 & - & 51 \\
$^{2}10_{1/2}[70,1_2^-]$ & 1947 $^{\ddagger}$ & 0& 0& 1 & 0 & 0 & 0 \\
$^{2}10_{3/2}[70,1_2^-]$ & 1947 $^{\ddagger}$ & 0 & 0 & 1 & 0 & 0& 0\\
$^{2}10_{5/2}[70,2_1^+]$ & 1859 $^{\ddagger}$ & 10& 0 & 97& 7 & - & 13 \\
$^{4}10_{3/2}[56,0_3^+]$ & 2103 & 0 & 0 & 0& 0 & 0 & 0 \\
\noalign{\smallskip}
\hline
\hline
\end{tabular}
\end{table}

\begin{table*}
\caption{As Table \ref{tab:missing-nuc}, but for missing $\Sigma$ resonances. } 
\label{tab:sigma-missing} 
\centering
\small
\begin{tabular}{ccccccccccccccccc}
\hline
\hline
\noalign{\smallskip}
$\Sigma$ & Mass & $N \overline{K}$ & $\Sigma \pi$ & $\Lambda \pi$ & $\Sigma \eta$ & $\Xi K$ & $\Delta \overline{K}$ & $\Sigma^* \pi$ & $\Sigma^* \eta$ & $\Xi^* K$ 
& $N \overline{K}^*$ & $\Sigma \rho$ & $\Lambda \rho$ & $\Sigma \omega$&$\Delta\overline{K}^*$ \\
\noalign{\smallskip}
\hline
\\
\multicolumn{16}{c}{$ $ $U(7)$ Model}\\
\multicolumn{16}{c}{$ $ \line(1,0){350}}\\

\noalign{\smallskip}
$^{4}8_{1/2}[70,1_1^-]$ & 1822 & 24 & 11 & 6 & { 20} & 10 & 5 & 4 & -- & -- & -- & -- & -- & -- & -- \\
$^{4}8_{3/2}[70,1_1^-]$ & 1822 & 22 & 4 & 8 & { 0} & 0 & 602 & 98 & -- & -- & -- & -- & -- & -- & -- \\
$^{2}8_{1/2}[70,0_1^+]$ & 1822 $^{\ddagger}$ & 1 & 20 & 1 & { 1} & 0 & 33 & 9 & -- & -- & -- & -- & -- & -- & -- \\
$^{2}8_{J}  [20,1_1^+]$ & 1849 $^{\ddagger}$ & 0 & 0 & 0 & 0 & 0 & 0 & 0 & -- & -- & 0 & -- & -- & --& --  \\
$^{2}8_{3/2}[56,2_1^+]$ & 1872 & 4 & 62 & 19 & { 23} & 12 & 17 & 6 & -- & -- & 12 & -- & -- & --& --  \\
$^{4}8_{3/2}[70,0_1^+]$ & 1926 & 0 & 0 & 0 & { 5} & 1 & 65 & 12 & -- & -- & 25 & -- & 3  & --& --  \\
$^{2}8_{3/2}[70,2_1^+]$ & 1999 & 1 & 31 & 1 & { 7} & 14 & 28 & 8 & { 1} & -- & 26 & 11 & 6  & { 1} & -- \\
$^{2}8_{5/2}[70,2_1^+]$ & 1999 & 4 & 76 & 6 & { 1} & 1 & 60 & 13 & { 4} & -- & 7  & 13 & 21 & { 3}& --  \\
$^{2}8_{5/2}[70,2_1^-]$ & 1999 & 0 & 0 & 0 & 0 & 0 & 0 & 0 & 0 & -- & 0  & 0  & 0  & 0 & --  \\
$^{4}8_{1/2}[70,2_1^+]$ & 2095 & 4 & 2 & 1 & { 3} & 4 & 18 & 4 & { 4} & 1 & 23 & 5  & 9  & { 7} & --  \\
$^{4}8_{3/2}[70,2_1^+]$ & 2095 & 2 & 1 & 1 & { 2} & 2 & 84 & 18 & { 13} & 2 & 39 & 7  & 14 & { 11}& --  \\
$^{4}8_{5/2}[70,2_1^+]$ & 2095 $^{\ddagger}$ & 15 & 3 & 5 & { 1} & 0 & 128 & 29 & { 18} & 3 & 45 & 8  & 15 & { 11}& --  \\
$^{4}8_{7/2}[70,2_1^+]$ & 2095 & 69 & 13 & 24 & { 2} & 1 & 54 & 15 & { 1} & 0 & 17 & 0  & 3  & { 1} & --  \\
$^{4}8_{J}  [70,2_1^-]$ & 2095 & 0 & 0 & 0 & 0 & 0 & 0 & 0 & 0 & 0 & 0  & 0  & 0  & 0& --   \\
$^{2}8_{1/2}[70,1_2^-]$ & 1957 $^{\ddagger}$ & 0 & 0 & 0 & 0 & 0 & 0 & 0 & 0 & -- & 0 & -- & 0 & --& --  \\
$^{2}8_{3/2}[70,1_2^-]$ & 1957 & 0 & 0 & 0 & 0 & 0 & 0 & 0 & 0 & -- & 0 & -- & 0 & --& --  \\
$^{4}8_{1/2}[70,1_2^-]$ & 2055 & 0 & 0 & 0 & 0 & 0 & 0 & 0 & 0 & -- & 0 & 0 & 0 & 0& --  \\
$^{4}8_{3/2}[70,1_2^-]$ & 2055 & 0 & 0 & 0 & 0 & 0 & 1 & 0 & 0 & -- & 0 & 0 & 0 & 0 & -- \\
$^{4}8_{5/2}[70,1_2^-]$ & 2055 & 0 & 0 & 0 & 0 & 0 & 0 & 0 & 0 & -- & 0 & 0 & 0 & 0& --  \\
\noalign{\smallskip}
\multicolumn{16}{c}{$ $ \line(1,0){350}}\\
\multicolumn{16}{c}{$ $ hQM}\\
\multicolumn{16}{c}{$ $ \line(1,0){350}}\\
\noalign{\smallskip}
$^{2}8_{ 3/2}[56,2_1^+]$& 1906 & 4 & 69 & 22 & 19 & 20 & 20 & 7 & -- & -- & 21 & -- & -- & -- & --  \\
$^{4}8_{ 1/2}[70,1_1^-]$ & 1914 & 9 & 7 & 3 & 17 & 22 & 22 & 9 & -- & -- & 100 & -- & -- & --& --  \\
$^{2}8_{ 1/2}[56,0_3^+]$ & 2050 $^\dagger$ & 0 & 0 & 0 & 0 & 0 & 0 & 0 & -- & -- & 0 & -- & -- & --& --  \\
$^{2}8_{ 3/2}[70,2_1^+]$ & 2072 & 1 & 31 & 1 & 7 & 16 & 39 & 11 & -- & -- & 29 & -- & -- & --& --  \\
$^{2}8_{ 5/2}[70,2_1^+]$ & 2072 & 5 & 89 & 7 & 3 & 3   & 66 & 15 & -- & -- & 11 & -- & -- & --& --  \\
$^{2}8_{ 1/2}[70,1_2^-]$ & 2149 & 0 & 0 & 0 & 0 & 0 & 0 & 0 & 0 & 0 & 0 & 0 & 0 & 0 & 0 \\ 
$^{4}8_{ 1/2}[70,2_1^+]$ & 2187 & 3 & 2 & 1 & 3 &4 & 16& 3 & 5 &1 & 21& 6 & 9 & 2 &16\\
$^{4}8_{ 3/2}[70,2_1^+]$ & 2187 & 1 & 1 & 1 & 1 &2 & 91& 21& 17&5 & 40& 10& 15& 3 &18\\ 
$^{4}8_{ 5/2}[70,2_1^+]$ & 2187 & 19& 3 & 6 & 1 & 0&147& 34& 23&6 & 52& 10& 18& 3  &49\\ 
$^{4}8_{ 7/2}[70,2_1^+]$ & 2187 & 83& 15& 29& 2 & 1&83 & 21& 3 &0 & 30& 2 & 7 & 1 &159\\ 
$^{2}8_{ J }[20,1_1^+]$ & 2238 & 0 & 0 & 0 & 0 & 0& 0 & 0 & 0 &0 & 0 & 0 & 0 & 0  &0 \\ 
$^{4}8_{ J}[70,1_2^-]$ & 2263 & 0 & 0 & 0 & 0 & 0 & 0 & 0 & 0 & 0 & 0 & 0 & 0 & 0  & 0  \\ 
\noalign{\smallskip}
\hline
\hline
\end{tabular}
\end{table*}

\begin{table*}
\caption{As Table \ref{tab:missing-nuc}, but for missing  $\Sigma^*$  resonances. } 
\label{tab:sigma*-missing} 
\centering
\small
\begin{tabular}{cccccccccccccccc}
\hline
\noalign{\smallskip}
$\Sigma$ & Mass & $N \overline{K}$ & $\Sigma \pi$ & $\Lambda \pi$ & $\Sigma \eta$ & $\Xi K$ & $\Delta \overline{K}$ & $\Sigma^* \pi$ & $\Sigma^* \eta$ & $\Xi^* K$ 
& $N \overline{K}^*$ & $\Sigma \rho$ & $\Lambda \rho$ & $\Sigma \omega$&$\Delta\overline{K}^*$ \\
\noalign{\smallskip}
\hline
\\
\multicolumn{16}{c}{$ $ $U(7)$ Model}\\
\multicolumn{16}{c}{$ $ \line(1,0){350}}\\

\noalign{\smallskip}
$^{2}10_{1/2}[70,1_1^-]$ & 1755 & 4 & 5 & 4 & 11 & -- & 1 & 30 & -- & -- & --  & --  & --  & -- & -- \\
$^{2}10_{3/2}[70,1_1^-]$ & 1755 & 9 & 6 & 14 & 0 & -- & 181 & 165 & -- & -- & --  & --  & --  & -- & -- \\
$^{2}10_{1/2}[70,0_1^+]$ & 1863 & 0 & 1 & 0 & 1 & 0 & 45 & 39 & -- & -- & 5  & -- & -- & -- & --\\
$^{4}10_{1/2}[56,2_1^+]$ & 2012 & 12 & 18 & 16 & 35 & 14 & 21 & 17 & 0 & -- & 24 & 6  & 28 & 76& --  \\
$^{4}10_{3/2}[56,2_1^+]$ & 2012 & 6 & 9 & 8 & 18 & 7 & 79 & 69 & 1 & -- & 35 & 9  & 40 & 106& -- \\
$^{4}10_{5/2}[56,2_1^+]$ & 2012 & 11 & 7 & 15 & 1 & 0 & 112 & 101 & 1 & -- & 37 & 9  & 41 & 106& -- \\
$^{2}10_{3/2}[70,2_1^+]$ & 2037 $^{\ddagger}$ & 1 & 1 & 1 & 2 & 1 & 38 & 42 & 0 & 0 & 28 & 12 & 36 & 315& -- \\
$^{2}10_{5/2}[70,2_1^+]$ & 2037 & 5 & 4 & 7 & 1 & 0 & 63 & 56 & 1 & 0 & 10 & 2  & 9  & 62& --  \\
$^{2}10_{J}  [70,2_1^-]$ & 2037 & 0 & 0 & 0 & 0 & 0 & 0 & 0 & 0 & 0 & 0  & 0  & 0  & 0 & -- \\
$^{4}10_{3/2}[56,0_2^+]$ & 1765 $^{\ddagger}$ & 8 & 8 & 9 & 1 & -- & 13 & 38 & -- & -- & --  & --  & --  & --& -- \\
$^{2}10_{J}  [70,1_2^-]$ & 1996 & 0 & 0 & 0 & 0 & 0 & 0 & 0 & 0 & -- & 0  & 0  & 0  & 0 & -- \\
\noalign{\smallskip}
\multicolumn{16}{c}{$ $ \line(1,0){350}}\\
\multicolumn{16}{c}{$ $ hQM}\\
\multicolumn{16}{c}{$ $ \line(1,0){350}}\\
\noalign{\smallskip}
$^{4}10_{ 3/2}[56,0_2^+]$ & 1883 & 6 & 9 & 8 & {11} & 2 & 68 & 63 & -- & -- & 14 & --  & --  & --& --   \\
$^{4}10_{ 3/2}[56,2_1^+]$ & 2085 & 6 & 9 & 9 & 18 & 8 & 86 & 76 & 2 & -- & 39 & 25 & 53 & 19& --  \\
$^{2}10_{ 3/2}[70,2_1^+]$& 2136 & 0 & 1 & 1 & 2 &1 &61 &63 & 0 &2 & 29&25 &43 &20&1 \\
$^{2}10_{ 5/2}[70,2_1^+]$& 2136 & 7 & 5 & 9 & 2 & 0& 70& 63& 1 &10& 17& 7 & 18& 5 &4 \\
$^{2}10_{ J}[70,1_2^-]$& 2212 & 0 & 0 & 0 & 0 & 0 & 0 & 0 & 0 & 0 & 0 & 0 & 0 & 0  & 0 \\
\noalign{\smallskip}
\hline
\hline
\end{tabular}
\end{table*}

\begin{table}
\caption{As Table \ref{tab:missing-nuc}, but for missing $\Lambda$ resonances.} 
\label{tab:lambda-missing} 
\centering
\small
\begin{tabular}{ccccccccccc}
\hline
\hline
\noalign{\smallskip}
$\Lambda$ & Mass & $N \overline{K}$ & $\Sigma \pi$ & $\Lambda \eta$ & $\Xi K$ & $\Sigma^* \pi$ & $\Xi^* K$ 
& $N \overline{K}^*$ & $\Sigma \rho$ & $\Lambda \omega$ \\
\noalign{\smallskip}
\hline
\\
\multicolumn{11}{c}{$ $$U(7)$ Model }\\
\multicolumn{11}{c}{$ $ \line(1,0){250}}\\
\noalign{\smallskip}
$^{4}8_{3/2}[70,1_1^-]$ & 1799 & 0 & 15 & { 1} & -- & 447 & -- & --  & --  & -- \\
$^{2}8_{J}  [20,1_1^+]$ & 1826 & 0 & 0 & 0 & 0 & 0 & -- & --  & --  & -- \\
$^{4}8_{3/2}[70,0_1^+]$ & 1904 & 0 & 3 & { 4} & 2 & 54 & -- & 0   & -- & { 0}  \\
$^{2}8_{3/2}[70,2_1^+]$ & 1978 & 27 & 6 & { 4} & 10 & 31 & -- & 56  & 1  & { 4}  \\
$^{2}8_{5/2}[70,2_1^+]$ & 1978 & 109 & 12 & { 2} & 1 & 58 & -- & 123 & 3  & { 16} \\
$^{2}8_{J}  [70,2_1^-]$ & 1978 & 0 & 0 & 0 & 0 & 0 & -- & 0   & 0  & 0  \\
$^{4}8_{J}  [70,2_1^-]$ & 2074 & 0 & 0 & 0 & 0 & 0 & 0 & 0   & 0  & 0  \\
$^{4}8_{1/2}[70,2_1^+]$ & 2075 & 0 & 13 & { 11} & 12 & 17 & 1 & 0   & 20 & { 9}  \\
$^{4}8_{3/2}[70,2_1^+]$ & 2075 & 0 & 6 & { 6} & 6 & 82 & 4 & 0   & 28 & { 13} \\
$^{4}8_{7/2}[70,2_1^+]$ & 2075 & 0 & 51 & { 10} & 2 & 57 & 0 & 0   & 1  & { 2} \\
$^{2}8_{J  }[70,1_2^-]$ & 1936 & 0 & 0 & 0 & 0 & 0 & -- & 0   & -- & 0 \\
$^{4}8_{1/2}[70,1_2^-]$ & 2034 & 0 & 0 & 0 & 0 & 0 &  0 & 0   & 0  & 0 \\
$^{4}8_{3/2}[70,1_2^-]$ & 2034 & 0 & 0 & 0 & 0 & 1 &  0 & 0   & 0  & 0 \\
$^{4}8_{5/2}[70,1_2^-]$ & 2034 & 0 & 0 & 0 & 0 & 0 &  0 & 0   & 0  & 0 \\
\noalign{\smallskip}
\multicolumn{11}{c}{$ $ \line(1,0){250}}\\
\multicolumn{11}{c}{$ $ hQM}\\
\multicolumn{11}{c}{$ $ \line(1,0){250}}\\
\noalign{\smallskip}
$^{4}8_{ 3/2}[70,1_1^-]$ & 1837 & 0 & 15 & { 2} & -- & 477 & -- & --  & --  & -- \\
$^{2}8_{ 3/2}[70,2_1^+]$& 1995 & 38 & 8 & { 0} & 10 & 29 & -- & 55  & 2 & { 4}  \\
$^{2}8_{1/2 }[70,1_2^-]$ & 2072 & 0 & 0 & 0 & 0 & 0 & -- & 0   & -- & 0 \\
$^{2}8_{3/2}[70,1_2^-]$  & 2072 & 0 & 0 & 0 & 0 & 0 & 0 & 0   & 0  & 0 \\
$^{4}8_{3/2}[70,0_1^+]$  & 2110 & 0 & 0 & {1} & 4 & 35 & 11 & 0 & 41 & 8 \\
$^{4}8_{1/2}[70,2_1^+]$  & 2110 & 0 & 18& 13& 12& 35& 2 & 0   & 23 & 9 \\
$^{4}8_{3/2}[70,2_1^+]$  & 2110 & 0 & 10& 6 & 2 & 87& 7 & 0   & 33 &14 \\
$^{4}8_{7/2}[70,2_1^+]$  & 2110 & 0 & 50&10 & 2 & 19& 0 & 0  & 2  & 2 \\
$^{2}8_{J  }[20,1_1^+]$  & 2160 & 0 & 0 & 0 & 0 & 0 & 0 & 0   & 0  & 0 \\
$^{4}8_{1/2}[70,1_2^-]$  & 2186 & 0 & 0 & 0 & 0 & 0 & 0 & 0 & 0 & 0 \\
$^{4}8_{3/2}[70,1_2^-]$  & 2186 & 0 & 0 & 0 & 0 & 1 & 0 & 0 & 0 & 0 \\
$^{4}8_{5/2}[70,1_2^-]$  & 2186 & 0 & 0 & 0 & 0 & 0 & 0 & 0 & 0 & 0 \\
\noalign{\smallskip}
\hline
\hline
\end{tabular}
\end{table}

\begin{table}
\caption{As Table \ref{tab:missing-nuc}, but for missing  $\Lambda^*$  resonances.} 
\label{tab:lambda*-missing} 
\centering
\small
\begin{tabular}{ccccccccccc}
\hline
\hline
\noalign{\smallskip}
$\Lambda$ & Mass & $N \overline{K}$ & $\Sigma \pi$ & $\Lambda \eta$ & $\Xi K$ & $\Sigma^* \pi$ & $\Xi^* K$ 
& $N \overline{K}^*$ & $\Sigma \rho$ & $\Lambda \omega$ \\
\noalign{\smallskip}
\hline
\\
\multicolumn{11}{c}{$ $ $U(7)$ Model}\\
\multicolumn{11}{c}{$ $ \line(1,0){250}}\\
\noalign{\smallskip}
$^{2}1_{1/2}[70,0_1^+]$ & 1756 & 29 & 44 & { 14} & -- & & & -- & -- & -- \\
$^{4}1_{J}  [20,1_1^+]$ & 1891 &  0 &  0 &  0 &  0 & & &  0 & -- & -- \\
$^{2}1_{3/2}[70,2_1^+]$ & 1939 & 35 & 66 & { 36} & 17 & & & 39 & -- &  { 6} \\
$^{2}1_{5/2}[70,2_1^+]$ & 1939 & 88 & 85 & { 10} &  0 & & & 94 & -- & { 15} \\
$^{2}1_{J}  [70,2_1^-]$ & 1939 &  0 &  0 &  0 &  0 & & &  0 & -- &  0 \\
$^{2}1_{J}  [70,1_2^-]$ & 1896 &  0 &  0 &  0 &  0 & & &  0 & -- &  0 \\
\noalign{\smallskip}
\multicolumn{11}{c}{$ $ \line(1,0){250}}\\
\multicolumn{11}{c}{$ $ hQM}\\
\multicolumn{11}{c}{$ $ \line(1,0){250}}\\
\noalign{\smallskip}
$^{2}1_{1/2}  [70,1_2^-]$ & 2008 &  0 &  1 &  0 &  0 &-- &-- &  1 & -- &  0 \\
$^{2}1_{3/2}  [70,1_2^-]$ & 2008 &  0 &  0 &  0 &  0 &-- &-- &  0 & -- &  0 \\
\noalign{\smallskip}
\hline
\hline
\end{tabular}
\end{table}

\begin{table*}
\caption{As Table \ref{tab:missing-nuc}, but for missing $\Xi$  resonances. } 
\label{tab:xi-missing} 
\begin{tabular}{cccccccccccc}
\hline
\hline
\noalign{\smallskip}
$\Xi$ & Mass & $\Sigma \overline{K}$ & $\Lambda \overline{K}$ & $\Xi \pi$ & $\Xi \eta$ 
& $\Sigma^* \overline{K}$ & $\Xi^* \pi$ & $\Lambda \overline{K}^*$&$\Sigma\overline{K}^*$& $\Xi\rho$&$\Xi\omega$ \\
\noalign{\smallskip}
\hline
\\
\multicolumn{12}{c}{$ $ $U(7)$ Model}\\
\multicolumn{12}{c}{$ $ \line(1,0){280}}\\
\noalign{\smallskip}
$^{2}8_{1/2}[70,0_1^+]$ & 1932 & 36 & 6 & 1 & { 11} & 7 & 13 & -- & --& --& --\\
$^{4}8_{1/2}[70,1_1^-]$ & 1932 & 43 & 20 & 69 & { 0} & 0 & 4 & --& --& -- & --\\
$^{4}8_{3/2}[70,1_1^-]$ & 1932 & 4 & 7 & 22 & { 0} & 216 & 152 & -- & --& --& --\\
$^{4}8_{5/2}[70,1_1^-]$ & 1932 & 23 & 39 & 132 & { 0} & 2 & 19 & -- & --& --& --\\
$^{2}8_{J}  [20,1_1^+]$ & 1957 & 0 & 0 & 0 & 0 & 0 & 0 & -- & --& --& --\\
$^{2}8_{3/2}[56,2_1^+]$ & 1979 & 198 & 7 & 6 & { 47} & 4 & 7 & --& --& --& -- \\
$^{2}8_{5/2}[56,2_1^+]$ & 1979 & 59 & 5 & 4 & { 1} & 20 & 27 & --& --& -- & --\\
$^{4}8_{3/2}[70,0_1^+]$ & 2031 & 2 & 1 & 3 & { 0} & 24 & 19 & 2& --& --& --  \\
$^{2}8_{1/2}[56,0_2^+]$ & 1727 & 26 & 4 & 3 & -- & -- & 2 & --& --& --& -- \\
\noalign{\smallskip}
\multicolumn{12}{c}{$ $ \line(1,0){280}}\\
\multicolumn{12}{c}{$ $ hQM}\\
\multicolumn{12}{c}{$ $ \line(1,0){280}}\\
\noalign{\smallskip}
$^{2} 8_{1/2}[56,0_2^+]$ & 1843 & 125 & 6   & 5   & --    & --    & 15 & --& --& --& -- \\
$^{4} 8_{3/2}[70,1_1^-]$  & 2053 & 8 & 11 & 37 & { 0} & 223 & 154 & -- & --  & --& --\\
$^{2} 8_{1/2}[56,0_3^+]$ & 2190 & 0 & 0 & 0 & 0 & 0 & 0  & 0 & 0 & 0 & 0  \\
$^{2} 8_{1/2}[70,1_2^-]$  & 2288 & 0 & 0 & 0 & 0 & 0 & 0 & 0 & 0 & 0 & 0 \\
$^{2} 8_{3/2}[70,1_2^-]$  & 2288 & 0 & 0 & 0 & 0 & 0 & 0 & 0 & 0 & 0 & 0 \\
$^{4} 8_{1/2}[70,2_1^+]$ & 2327 & 3 & 1 & 8 & 1 & 6   & 5  & 8  & 10  & 40 & 1    \\
$^{4} 8_{3/2}[70,2_1^+]$ & 2327 & 2 & 1 & 4 & 0 & 35  & 32 & 16 & 16  & 62 & 1   \\
$^{4} 8_{5/2}[70,2_1^+]$ & 2327 & 6 & 7 & 24& 0 & 57  & 53 & 20 & 18  & 69 & 1   \\
$^{4} 8_{7/2}[70,2_1^+]$ & 2327 & 26& 33&108& 1 & 33  & 33 & 11 & 5   & 16 & 0   \\
$^{2} 8_{J  }[20,1_1^+]$  & 2377 & 0 & 0 & 0 & 0 & 0    & 0 & 0 & 0 &0 & 0   \\ 
$^{4} 8_{J}[70,1_2^-]$  & 2403  & 0 & 0 & 0 & 0 & 0 & 0 & 0 & 0 & 0 & 0   \\
\noalign{\smallskip}
\hline
\hline
\end{tabular}
\end{table*}

\begin{table*}
\caption{As Table \ref{tab:missing-nuc}, but for missing  $\Xi^*$ resonances. } 
\label{tab:xi*-missing} 
\begin{tabular}{cccccccccccc}
\hline
\hline
\noalign{\smallskip}
$\Xi$ & Mass & $\Sigma \overline{K}$ & $\Lambda \overline{K}$ & $\Xi \pi$ & $\Xi \eta$ 
& $\Sigma^* \overline{K}$ & $\Xi^* \pi$ & $\Lambda \overline{K}^*$&$\Sigma\overline{K}^*$& $\Xi\rho$&$\Xi\omega$ \\
\noalign{\smallskip}
\hline
\\
\multicolumn{12}{c}{$ $ $U(7)$ Model}\\
\multicolumn{12}{c}{$ $ \line(1,0){280}}\\
\noalign{\smallskip}
$^{2}10_{1/2}[70,1_1^-]$ & 1869 & 17 & 10 & 7 & { 7} & -- & 7 & --& --& -- & -- \\
$^{2}10_{3/2}[70,1_1^-]$ & 1869 & 5 & 10 & 9 & { 0} & -- & 61 & --& --& --& --  \\
$^{2}10_{1/2}[70,0_1^+]$ & 1971 & 2 & 1 & 1 & { 2} & 51 & 14 & -- & --& --& -- \\
$^{4}10_{3/2}[56,0_2^+]$ & 1878 & 19 & 16 & 13 & { 1} & -- & 12 & --& --& --& --  \\ 
\noalign{\smallskip}
\multicolumn{12}{c}{$ $ \line(1,0){280}}\\
\multicolumn{12}{c}{$ $ hQM}\\
\multicolumn{12}{c}{$ $ \line(1,0){280}}\\
\noalign{\smallskip}
$^{4}10_{3/2}[56,0_2^+]$ & 2022 & 19 & 12 & 13 & {12} & -- & 24 & --&--&--&-- \\ 
$^{4}10_{1/2}[56,2_1^+]$ & 2225 & 33& 19& 23& 35& 33  & 7  & 40   & 34  &31  & 16   \\
$^{4}10_{3/2}[56,2_1^+]$ & 2225 & 17& 10& 12& 17& 133 & 29 & 61   & 50  & 44 & 23   \\
$^{4}10_{5/2}[56,2_1^+]$ & 2225 & 14& 19& 16& 3 & 194 & 44 & 66    & 51  & 45 & 24    \\
$^{4}10_{7/2}[56,2_1^+]$ & 2225 & 64& 87& 70& 13& 56  & 16 & 13      & 3   & 2  & 1  \\
$^{2}10_{1/2}[70,1_2^-]$ & 2352  & 0 & 0 & 0 & 0 & 0 & 0 & 0 & 0 & 0 & 0 \\
$^{2}10_{3/2}[70,1_2^-]$ & 2352  & 0 & 0 & 0 & 0 & 1 & 0 & 0 & 0 & 0 & 0 \\
$^{4}10_{3/2}[56,0_3^+]$ & 2369 & 0 & 0 & 0 & 0 & 0 & 0 & 0 & 0 & 0 & 0  \\
\noalign{\smallskip}
\hline
\hline
\end{tabular}
\end{table*}
\begin{table}
\caption{As Table \ref{tab:missing-nuc}, but for missing $\Omega$ resonances. } 
\label{tab:omega}
\begin{tabular}{cccccc}
\hline
\hline
\noalign{\smallskip}
$\Omega$ & Mass & $\Xi \overline{K}$ & $\Xi^* \overline{K}$&$\Omega \eta$& $\Xi\overline{K}^*$ \\
\noalign{\smallskip}
\hline
\\
\multicolumn{6}{c}{$ $ $U(7)$ Model}\\
\multicolumn{6}{c}{$ $ \line(1,0){170}}\\
\noalign{\smallskip}
$^{2}10_{1/2}[70,1_1^-]$ & 1989 & 68 & -- & --& --\\
$^{2}10_{3/2}[70,1_1^-]$ & 1989 & 20 & --& --& -- \\
$^{2}10_{1/2}[70,0_1^+]$ & 2085 &  8 & 32& --& -- \\
$^{4}10_{3/2}[56,0_2^+]$ & 1998 & 79 & -- & --& --\\
\noalign{\smallskip}
\multicolumn{6}{c}{$ $ \line(1,0){170}}\\
\multicolumn{6}{c}{$ $ hQM}\\
\multicolumn{6}{c}{$ $ \line(1,0){170}}\\
\noalign{\smallskip}
$^{2}10_{1/2}[70,1_1^-]$ & 2142 & 26 & 48 & -- &--\\
$^{2}10_{3/2}[70,1_1^-]$ & 2142 & 68 & 403 & --&-- \\
$^{4}10_{3/2}[56,0_2^+]$ & 2162 & 68 & 102 & -- &-- \\
$^{4}10_{1/2}[56,2_2^+]$ & 2364 &109 & 34 & 27 &155\\
$^{4}10_{3/2}[56,2_2^+]$ & 2364 &  55 &137 & 88 &225\\
$^{4}10_{5/2}[56,2_2^+]$ & 2364 &  69 & 199 &117 &234\\
$^{4}10_{7/2}[56,2_2^+]$ & 2364 & 308 & 58 & 4 &23 \\
$^{2}10_{1/2}[70,1_2^-]$ & 2492 & 0 & 0 & 0 & 0 \\
$^{2}10_{3/2}[70,1_2^-]$ & 2492 & 0 & 1 & 0 & 0 \\
$^{4}10_{3/2}[56,0_3^+]$ & 2508 & 0 & 0 & 0 & 0 \\
\noalign{\smallskip}
\hline

\hline
\end{tabular}
\end{table}

\subsection{Open-flavor strong decays calculated by using the hQM}
\label{Open-flavor strong decays calculated using the hQM model spectrum}
Below, we provide results for the open-flavor decay widths of strange and non-strange baryons into light baryons plus pseudoscalar or vector mesons in the $^3P_0$ model formalism of Sec. \ref{3P0 model}, using the hQM results of Refs. \cite{pl,chin,hQM-strange}. For the results, see Tables \ref{tab:nuc}-\ref{tab:delta-missing}.
The decays are calculated with the new values of the $^3P_0$ model parameters of Table \ref{tab:parameters-3P0+hQM}, which we fitted to a sample of 9 transitions: $\Delta \rightarrow N \pi$, $N(1520) \rightarrow N \pi$, $N(1535) \rightarrow N \pi$, $N(1650) \rightarrow N \pi$, $N(1680) \rightarrow N \pi$, $N(1720) \rightarrow N \pi$, $\Delta(1905) \rightarrow N \pi$, $\Delta(1910) \rightarrow N \pi$ and $\Delta(1920) \rightarrow N \pi$.

Although we use the same decay model as in Sec. \ref{Open-flavor strong decays calculated using the U(7) model spectrum}, there are some differences with respect from the previous case: 1) As in Sec. \ref{Open-flavor strong decays calculated using the U(7) model spectrum}, for *** and **** states we use the experimental values of the masses, but the quantum number assignments are provided by the hQM and do not always coincide with those of the U(7) model, since the two models are different.
For example, this happens for the $N(1650)S_{11}$, $N(1700)D_{13}$ and $N(1875)D_{13}$. Thus, in these cases, we expect to obtain quite different results for the decay widths; 2) The hQM and U(7) models predict the existence of a few missing states below the energy of 2.1 GeV. The masses, quantum numbers and also quantity of missing states in the two previous models are different. Information on missing states is important to the experimentalists in their search for new baryon resonances.

\begin{table} 
\caption{Parameter values used in the calculations, in combination with the relativistic phase space factor of Eq. (\ref{eqn:rel-PSF}). The parameter values are fitted to a sample of 9 transitions: $\Delta \rightarrow N \pi$, $N(1520) \rightarrow N \pi$, $N(1535) \rightarrow N \pi$, $N(1650) \rightarrow N \pi$, $N(1680) \rightarrow N \pi$, $N(1720) \rightarrow N \pi$, $\Delta(1905) \rightarrow N \pi$, $\Delta(1910) \rightarrow N \pi$ and $\Delta(1920) \rightarrow N \pi$. The quantum number assignments for the decaying states are now taken from the hQM results of Ref. \cite{pl,chin} and Table \ref{tab:nuc}.}
\label{tab:parameters-3P0+hQM}  
\centering
\begin{tabular}{cc} 
\hline 
\hline 
\noalign{\smallskip}
Parameter           & Value \\ 
\noalign{\smallskip}
\hline 
\noalign{\smallskip}
$\gamma_0$   &  13.319 \\  
$\alpha_{b}$ & 2.758  \\ 
$\alpha_{c}$ &  2.454  \\ 
$\alpha_{d}$ & 0  \\ 
$m_n$        & 0.33  \\
$m_s$        & 0.55  \\
\noalign{\smallskip}
\hline 
\hline
\end{tabular}
\end{table}

\subsection{Comparison with other QM calculations}

The quality of our results is comparable to that of Refs. \cite{Capstick:1992th,Capstick:1998md}. 
Capstick and Roberts (CR) studied the strong decays of non-strange baryons, nucleon and delta resonances, by using Capstick and Isgur's relativized baryon model \cite{Capstick:1986bm} to describe the masses of unknown resonances, harmonic oscillator wave functions and an effective phase space \cite{Capstick:1992th,Capstick:1998md}; they did not calculate the open-flavor decays of strange baryons.
Our study is more complete, as it includes many more decay channels (such as, for example, decays into $\Xi$ + meson) and a detailed analysis of the decays of strange baryons.
Both calculations, ours and Capstick and Roberts' \cite{Capstick:1992th,Capstick:1998md}, are performed within the $^3P_0$ model, though there are some differences. 
The main one is in the pair-creation mechanism: in Refs.~\cite{Capstick:1992th,Capstick:1998md} it does not depend on the flavor of the created $q\overline{q}$ pair, 
while in the present case the strange quark pair-creation is suppressed with respect to the nonstrange pairs. The effects of this strangeness-suppression mechanism cannot be re-absorbed in a redefinition of the model parameters or in a different choice of the $^3P_0$ model vertex factor.
Whether $s \bar s$ pair-creation has to be suppressed or not may be evaluated by comparing theoretical predictions and experimental data for these particular decay channels.
Unfortunately, we think that the large uncertainties on experimental decay widths into channels due to $s \bar s$ pair-creation do not permit a strong definitive conclusion to be drawn.
Nevertheless, it is worthwhile to observe that the introduction of a strangeness suppression mechanism may be justified by experimental results concerning the electro-production ratios of $\Lambda K^+$, $\Sigma^* K$, $p \pi^0$ and $n \pi^+$ baryon meson-states from $N^*$'s \cite{Santopinto:2016fgs}.

In our paper, we also compared results obtained with relativistic or effective phase space factors. Our conclusion is that the quality of the results for the amplitudes calculated by using a phase space or the other are similar. Thus, we think that it is preferable to use a relativistic phase space, in order to reduce the number of unnecessary parametrizations.
Finally, we can say that both studies are characterized by the same problems (a few results are far from data, like $\Gamma_{N(1700) D_{13} \rightarrow \Delta \pi}$) and strengths (a great number of data is fitted with a few parameters); among other things, CR obtained for the decay widths $N(1535)S_{11} \rightarrow N \pi$, $N(1700)D_{13} \rightarrow \Delta \pi$ and $N(1720)P_{13} \rightarrow N \rho$, the following theoretical results 216 MeV, 778 MeV, and 11 MeV, respectively, which are to be compared with the experimental data $44-96$ MeV, $10-225$ MeV and $105-340$ MeV, respectively. 

We can also compare our results with those of Refs. \cite{Bijker:1996tr,Bijker:1994yr}. 
Bijker, Iachello and Leviatan (BIL) \cite{Bijker:1996tr,Bijker:1994yr} computed the open-flavor decay amplitudes within a modified version of the elementary meson emission model (EME), with two parameters, and used the U(7) algebraic model to calculate the baryon spectrum.
The EME is an effective model, in which the decay occurs by the emission of a meson from the decaying hadron. 
Even though the EME and $^3P_0$ models share some features, there are important differences. For example, the internal quark dynamics is invisible to the EME vertex, because it does not depend on the meson internal wave function, which brings a non-local character to the $^3P_0$ matrix elements. Moreover, as in the CR case, BIL did not consider a suppression of $s \bar {s}$ pair-creation, though it has been shown that this mechanism can be beneficial in meson strong decays \cite{Ferretti:2013faa,Ferretti:2013vua,bottomonium}.
It can also be shown that it is impossible to get a correspondence between EME  and $^3P_0$ models, unless the factor in front of the recoil term in the EME is taken equal to $k_0/2m = 1$. However, this 
factor has been taken as a free parameter in BIL (and also in other EME studies, Refs. \cite{Koniuk:1979vy,Bijker:1996tr,Bijker:1994yr}) and by fit it came out  to be equal 0.04, which means that the recoil term is practically absent in BIL. 
This fact, in combination with the non local character of the $^3P_0$ model, the quark form factor and flavor suppression, explain the differences in the two model predictions.  
A more detailed explanation is contained in Appendix \ref{EME vs 3P0}. 
Finally, we can say that the general quality of our results is comparable to those of Refs. \cite{Bijker:1996tr,Bijker:1994yr}. Both of them reproduce the general trend of the data, but, in some cases, they show a few large disagreement with  experiments. For example, the results of Refs. \cite{Bijker:1996tr,Bijker:1994yr} do not agree with the data in the case of the decays $N(1720)P_{13} \rightarrow \Lambda K$ and $N(1440)P_{11} \rightarrow \Delta \pi$, where they get null amplitudes, while ours do agree. On the contrary, our results for other channels, like $N(1520) D_{13} \rightarrow \Delta \pi$ and $\Lambda(1670) S_{01} \rightarrow N \bar K$, do not agree with the data, being much larger, while those of Refs.~\cite{Bijker:1996tr,Bijker:1994yr} do agree or they are closer. 

There was also an attempt to improve the study of the strong decays using covariant calculations with relativistic constituent quark models (rCQM) by Melde {\it et al.} \cite{MeldePRC,MeldePRD}. The authors computed the transitions for all $\pi$, $\eta$, and $K$ strong decay modes of several well-established non-strange and strange baryon states, using the so-called point-form spectator model (PFSM), whose non relativistic limit is the classic EME; they did not calculate decay amplitudes into baryon-vector meson pairs. 
Unfortunately, these results still cannot provide a satisfactory explanation of the experimental decay widths and, in general, underestimate the available experimental data.
Finally, it is worthwhile to cite the results of a dynamical coupled-channels study of $\pi N \rightarrow \pi \pi N$ reactions of Ref. \cite{Kamano:2008gr}, where the authors provided a comprehensive analysis of world data of $\pi N$, $\gamma N$ and $N(e,e')$ reactions, including a coupled channel model for meson production reactions considering all possible final states. 

\subsection{Exotic states and alternative decay modes}

As widely discussed in the literature, there are several baryons (mesons) whose nature may not be a pure three-quark (quark-antiquark) one. Some well-known examples are the $X(3872)$ \cite{Kalashnikova:2005ui,Danilkin:2010cc,Ferretti:2013faa,Ferretti:2014xqa,Swanson:2003tb,Aceti:2012cb,Pennington:2007xr} and the $D_{s0}^*(2317)$ and $D_{s1}(2460)$ \cite{Hwang:2004cd,Ortega:2016mms} mesons.
Some alternative hypotheses (hadron-hadron molecules, hybrids, tetra/penta-quarks, and so on) might explain why na\"ive quark models fail to reproduce some of the main properties of these resonances, including mass and decay modes.

Let us focus on the $\Lambda^*(1405)$. By combining U(7) and $^3P_0$ models, we get a mass of 1641 MeV and a $\Sigma \pi$ amplitude of 230 MeV; these numbers can be compared with experimental results, namely $1402-1410$ MeV (mass) and $48-52$ MeV ($\Sigma \pi$ amplitude) \cite{Nakamura:2010zzi}. 
Such a strong deviation between theoretical predictions and data might be explained if we interpret the $\Lambda^*(1405)$ as a baryon-meson molecular state (for example, see Ref. \cite{Ramos:2005fe,Hall}). Given this, our predictions would refer to the lowest-lying $qqq$ state with the same quantum numbers as the $\Lambda^*(1405)$.

The $\Delta(1930)D_{35}$ represents another failure of our $^3P_0$/QM predictions, as we get a null $N \pi$ width, while it should be in the range $11-75$ MeV \cite{Nakamura:2010zzi}.
Nevertheless, unlike the $\Lambda^*(1405)$ case, here we obtain a prediction for the mass which is compatible with the experimental data \cite{Nakamura:2010zzi}. Possible explanations of this deviation from the data include the eventuality that the $\Delta(1930)D_{35}$ is a $\Delta \rho$ bound state \cite{Gonzalez:2008pv} or simply the fact that the $N \pi$ decay may proceed in a different way. It would be thus worthwhile to investigate whether the inclusion of baryon-meson higher Fock components in $\Delta(1930)D_{35}$'s wave function via the unquenched quark model (UQM) formalism \cite{uqm} may help to solve this problem.
Similar issues also occur in the $\Sigma(1750)$ and $\Sigma(1940)$ cases. See Refs. \cite{Oset:2009vf,Sarkar:2004jh,Sarkar:2009kx}.

Another interesting case of departure from experimental data is that of the $\Delta(1700)D_{33}$ $\Delta \pi$ width \cite{Nakamura:2010zzi}, for which we predict a larger value. On the contrary, in the molecular picture this decay mode is dynamically suppressed. It would be worthwhile to investigate this case in the UQM formalism and see if the introduction of continuum components, also determining a renormalization of the $\Delta(1700)$ wave function, may improve the quality of our result.
It would also be interesting to calculate the couplings for the $\Delta \eta$ virtual channel, which is relevant to several reactions. This coupling has been evaluated in the molecular model in Refs. \cite{Doring:2005bx,Nakabayashi:2006ut,Ajaka:2008zz}.

In addition to a calculation of the decay amplitudes within the UQM, it would also be interesting to investigate baryon-meson-meson decays. A possible way to do that is via the formalism of quasi-two-body decays, discussed in Refs. \cite{Ferretti:2014xqa,Kokoski:1985is,Capstick:1992th}. In quasi-two-body decays, the decay of a baryon $A$ into a baryon $B$ and mesons $C_1$ and $C_2$ proceeds as $A \rightarrow B^* C_1 \rightarrow B C_1 C_2$, where $B^*$ is a baryon resonance. Alternately, one may also decide to use the coupled-channel approach. Of particular interest is the $N(1710)D_{13} \rightarrow N \pi \pi$ decay mode, which is the main one of the $N(1710)$ ($40-90 \%$); it is worthwhile to note that this is larger than the $N \pi$ width, even if the latter has more phase space for the decay. For example, see Refs. \cite{Khemchandani:2008rk}.

\section{Summary and conclusion}
\label{Discussion of the results}
We computed the open-flavor strong decays of light baryons (i.e. made up of $u$, $d$, $s$ valence quarks) into baryon-pseudoscalar and baryon-vector mesons using a modified version of the $^3P_0$ pair-creation model \cite{Micu,LeYaouanc}, in which we considered a flavor-dependent pair-creation strength to suppress the contributions from heavier $q \bar{q}$ pairs, like $s \bar s$ with respect to $u \bar u$ ($d \bar d$). 

The baryon models, which we used in our study to get predictions for missing or higher-lying states, were the $U(7)$ \cite{Bijker:1994yr} and hypercentral \cite{pl,chin} models. 
The possibility of using two models to extract the baryon spectrum makes it possible to give two different points of view, especially in the study of the energy region above $1.8-2$ GeV and the related problem of the missing resonances. Indeed, an important difference between the U(7) and hQM models is in the number of missing states that they predict and also in the quantum number predictions for some *** and **** states, like the $N(1875)D_{13}$. In a subsequent paper \cite{subsek}, the present results will be extended up to to an energy region (2.5 GeV) which will be tested by forthcoming experiments at the JLab. 

It is worthwhile to enumerate some of the difficulties and problems connected to this type of calculations. 
One problem is related to the difficulty of assigning quantum numbers to resonances within a QM. 
Sometimes, this can generate strong conflicts between theoretical results and experimental data.
For example, this is the case of the $N(1875)D_{13}$, whose decay amplitudes change significantly whether we use the $^3P_0$ model in combination with the U(7) or hQM models. See Tables \ref{tab:nuc}..
Another problem has to do with the large quantity of decay thresholds, sometimes lying at similar energies or almost overlapping with one another. 
Thus, we think that a more complete study would require the introduction of continuum coupling effects (i.e. higher Fock components) in the baryon wave functions. 
In some cases, the presence of a threshold can deeply influence the quark structure of a hadron, close in energy, as in the well-known case of the $X(3872)$ meson \cite{Ferretti:2013faa,Ferretti:2014xqa,Pennington:2007xr,Danilkin:2010cc}.
In this respect, it is worthwhile to cite the results of the EBAC project, developed by Matsuyama $et$ $al.$ \cite{Matsuyama:2006rp}. This is a dynamical coupled-channel model for investigating the nucleon resonances in the meson production reactions induced by pions and photons.
See also the interesting coupled-channel model results of the Bonn-J\"ulich group of Refs. \cite{Doring}. 
A similar formalism, which would make it possible to include meson cloud effects in baryon and meson open- and hidden-flavor decays, will be the subject of a subsequent paper \cite{subsek}.

Finally, one of the most important points is related to the problem of the missing resonances. 
Is this a matter of degrees of freedom? In this case, the use of other types of models, characterized by a smaller number of effective degrees of freedoms, such as the quark-diquark one, could, at least partially, solve the problem \cite{Santopinto:2004hw}. Otherwise, does it have to do with the coupling of these missing states with other types of decay channels, more difficult to observe? This is still an open question.
Thus, we think that it would be worthwhile to compare the results for spectrum and decays of a three quark QM to those of other type of models, such as the quark-diquark one. Moreover, in the case of higher lying states, it would also be interesting to compare the predictions of three quark models to those for hybrid baryons, where baryons are described as bound states of three constituent quarks and a constituent gluon \cite{Guo:2007sm}.

\appendix

\section{Pair-creation vertex}
\label{Pair-creation vertex}
Analogously to what is done in Ref. \cite{Bonnaz:1999zj}, we can study different forms for the pair-creation vertex, to improve the description of the experimental data. The determination of the best vertex results from a $\chi^2$-analysis based on a sample of 9 transitions: $\Delta \rightarrow N \pi$, $N(1520) \rightarrow N \pi$, $N(1535) \rightarrow N \pi$, $N(1650) \rightarrow N \pi$, $N(1680) \rightarrow N \pi$, $N(1720) \rightarrow N \pi$, $\Delta(1905) \rightarrow N \pi$, $\Delta(1910) \rightarrow N \pi$ and $\Delta(1920) \rightarrow N \pi$. The different forms we consider are given by
\begin{eqnarray}
V_1(2p) &=& e^{-\alpha_d^2 p^2/2}  
\nonumber\\
V_2(2p) &=& (1 + \gamma_1 p^2) \, e^{-\alpha_d^2 p^2/2} 
\nonumber\\
V_3(2p) &=& 1 + \gamma_1 \, e^{-\alpha_d^2 p^2/2} 
\nonumber\\
V_4(2p) &=& 1 + (\gamma_1 + \gamma_2 p^2) \, e^{-\alpha_d^2 p^2/2}  
\label{tab:eff-vertex}
\end{eqnarray}
where $p^2=(\vec{p}_4 - \vec{p}_5)^2/4$.  
As observed in Ref. \cite{Bonnaz:1999zj}, the forms containing a $p_0$ parameter, such as $V \propto e^{-\alpha_d^2 (p-p_0)^2}$ or $1/[(p-p_0)^2+B]$, present a bump around $p - p_0$, and thus do not show the expected decreasing behavior. Thus, we do not include them in our analysis.
The quality of the description of the experimental data provided by the four vertices is equivalent (see Table \ref{tab:9transitions}). Thus, we choose the vertex with the smallest number of free parameters, the first vertex with
\begin{equation}
V(\vec{p}_4-\vec{p}_5) = e^{-\alpha_d^2 (\vec{p}_4 - \vec{p}_5)^2/8} \mbox{ }.
\label{gamma0eff}
\end{equation}
This is the one used in the calculations of Sec. \ref{Strong decay widths} and for the analytic derivation of the $^3P_0$ amplitudes of App. \ref{Ap1}.
The reader may object that we did not consider in our analysis the simplest choice for the pair-creation vertex, namely $V = 1$. Actually, this particular choice is a special case of $V_1(2p)$, i.e. when $\alpha_d = 0$.

\begin{table}
\caption{Comparison of the results obtained with different vertex functions, fitted to a selected number of experimental strong decays \cite{Nakamura:2010zzi}. Columns $2-5$ show the theoretical open-flavor decay widths, calculated with the vertices $V_i$ of Eq. (\ref{tab:eff-vertex}) in combination with the effective phase space factor of Eq. (\ref{eqn:eff-PSF}).}
\label{tab:9transitions} 
\centering
\begin{tabular}{cccccc}
\hline
\hline
\noalign{\smallskip}
Channel & $V_1$ & $V_2$ & $V_3$ & $V_4$ & Exp (MeV) \\
\noalign{\smallskip}
\hline
\noalign{\smallskip}
$\Delta(1232) \rightarrow N \pi$       & 115 & 118 & 116 & 120 & $114-120$ \\
$N(1520) \rightarrow N \pi$      & 102 & 98  & 101 & 98  & $ 55- 81$ \\
$N(1535) \rightarrow N \pi$      & 106 & 108 & 102 & 107 & $ 44- 96$ \\ 
$N(1650) \rightarrow N \pi$      & 71  & 72  & 68  & 72  & $ 60-162$ \\ 
$N(1680) \rightarrow N \pi$      & 63  & 55  & 60  & 50  & $ 78- 98$ \\ 
$N(1720) \rightarrow N \pi$      & 123 & 114 & 114 & 118 & $ 12- 56$ \\ 
$\Delta(1905) \rightarrow N \pi$ & 14  & 14  & 14  & 14  & $ 24- 60$ \\ 
$\Delta(1910) \rightarrow N \pi$ & 39  & 42  & 38  & 43  & $ 33-102$ \\
$\Delta(1920) \rightarrow N \pi$ & 14  & 16  & 14  & 16  & $  9- 60$ \\
\noalign{\smallskip}
\hline
\hline
\end{tabular}
\end{table}

\section{Spin Wave Functions}
In the following, we list the conventions used for the spin wave functions \cite{Koniuk:1979vy}:
\begin{equation}
\begin{array}{clrcl}
S=1/2 &: & |\chi^{\rho}_{1/2} \rangle & = & \frac{1}{\sqrt 2}
( |\uparrow \downarrow \uparrow \rangle
- |\downarrow \uparrow \uparrow \rangle) ~, \\ && \\
&& |\chi^{\lambda}_{1/2} \rangle & = & \frac{1}{\sqrt 6}
(2|\uparrow \uparrow \downarrow \rangle
- |\uparrow \downarrow \uparrow \rangle \\
& & & & - |\downarrow \uparrow \uparrow \rangle) ~, \\ && \\
S=3/2 & : & |\chi^{\rm S}_{3/2} \rangle & = & |\uparrow \uparrow \uparrow \rangle ~.
\end{array}
\end{equation}
We only show the state with the largest component of the projection $M_{\rm S}=S$. The other states are obtained by applying the lowering operator
in spin space.

\section{Flavor Wave Fuctions} 
\label{flavormeson}
The meson and baryon states are written according to the usual prescriptions.  
Below, we list the conventions used for the flavor wave functions of mesons and baryons.

\subsection*{Mesons}
Since the mixing angle $\theta_{\eta \eta'}$ between $\eta$ and $\eta'$ is small, we take $\theta_{\eta \eta'} =0$. Thus, we identify $\eta=\eta_8$ and $\eta'=\eta_1$.
\begin{itemize}
 \item {\bf The octet mesons}
 \begin{eqnarray}
|\pi^+\rangle&=& -|u\bar{d}\rangle \nonumber\\
|\pi^0\rangle&=&\frac{1}{\sqrt{2}}[|u\bar{u}\rangle-|d\bar{d}\rangle]\nonumber\\ \nonumber
|\pi^-\rangle&=&|d\bar{u}\rangle \\\nonumber
|K^+\rangle &=& -|u\bar{s}\rangle\\
|K^-\rangle &=& |s\bar{u}\rangle\\\nonumber
|K^0\rangle&=&- |d\bar{s}\rangle\\\nonumber
|\bar{K}^0\rangle&=& -|s\bar{d}\rangle\\\nonumber
|\eta \rangle&=& \frac{1}{\sqrt{6}} [ |u\bar{u}\rangle +|d\bar{d}\rangle- 2|s\bar{s}\rangle ]
\end{eqnarray} 
\item {\bf The singlet mesons}
 \begin{eqnarray}
  |\eta' \rangle&=& \frac{1}{\sqrt{3}}[ |u\bar{u}\rangle +|d\bar{d}\rangle+|s\bar{s}\rangle ]
 \end{eqnarray}
\end{itemize}

\subsection*{Baryons}
For the baryon flavor wave functions, $|(p,q),I,M_I,Y \rangle$, we adopt the convention of Ref. \cite{DeSwart}. We only show the highest charge state $M_I=I$ with $Q=I+Y/2$. The other charge states are obtained by applying the lowering operator in isospin space.
\begin{itemize}
\item {\bf The octet baryons} 
\begin{equation}
	\begin{array}{rcccl}
	|(1,1),\frac{1}{2},\frac{1}{2},1 \rangle & : & \phi_{\rho}(p) & = & \frac{1}{\sqrt 2} [ |udu \rangle - |duu \rangle ] \\
	& : & \phi_{\lambda}(p) & = & \frac{1}{\sqrt 6} [ 2|uud \rangle - |udu \rangle \\ 
	& & & & - |duu \rangle ]
	\end{array}
\end{equation}
\begin{equation}
	\begin{array}{rcccl}
	|(1,1),1,1,0 \rangle & : & \phi_{\rho}(\Sigma^+) & = & \frac{1}{\sqrt 2} [ |suu \rangle - |usu \rangle ]  \\
	& : & \phi_{\lambda}(\Sigma^+) & = &  \frac{1}{\sqrt 6} [ |suu \rangle + |usu \rangle \\
	& & & & - 2|uus \rangle ] 
	\end{array}
\end{equation}
\begin{equation}
	\begin{array}{rcccl}
	|(1,1),0,0,0 \rangle & : & \phi_{\rho}(\Lambda) & = & \frac{1}{\sqrt{12}} [ 2|uds \rangle - 2|dus \rangle \\
	& & & & - |dsu \rangle + |sdu \rangle  \\
	& & & & - |sud \rangle + |usd \rangle ] \\
	& : & \phi_{\lambda}(\Lambda) & = & \frac{1}{2} [- |dsu \rangle - |sdu \rangle \\
	& & & & + |sud \rangle + |usd \rangle ]
	\end{array}
\end{equation}
\begin{equation}
	\begin{array}{rcccl}
	|(1,1),\frac{1}{2},\frac{1}{2},-1 \rangle & : & \phi_{\rho}(\Xi^0) & = & \frac{1}{\sqrt 2} [ |sus \rangle - |uss \rangle ]  \\
	& : & \phi_{\lambda}(\Xi^0) & = & \frac{1}{\sqrt 6} [ 2|ssu \rangle \\ 
	& & & & - |sus \rangle - |uss \rangle ] 
	\end{array}
\end{equation}
\item {\bf The decuplet baryons}
\begin{equation}
	\begin{array}{rcccl}
	|(3,0),\frac{3}{2},\frac{3}{2},1 \rangle & : & \phi_S(\Delta^{++}) & = & |uuu \rangle
	\end{array}
\end{equation}
\begin{equation}
	\begin{array}{rcccl}
	|(3,0),1,1,0 \rangle & : & \phi_S(\Sigma^{+}) & = & \frac{1}{\sqrt 3} [ |suu \rangle + |usu \rangle \\
	& & & & + |uus \rangle ]
	\end{array}
\end{equation}
\begin{equation}
	\begin{array}{rcccl}
	|(3,0),\frac{1}{2},\frac{1}{2},-1 \rangle & : & \phi_S(\Xi^{0}) & = & \frac{1}{\sqrt 3} [ |ssu \rangle + |sus \rangle \\
	& & & & + |uss \rangle ]
	\end{array}
\end{equation}
\begin{equation}
	\begin{array}{rcccl}
	|(3,0),0,0,-2 \rangle & : & \phi_S(\Omega^{-}) & = &  |sss \rangle 
	\end{array}
\end{equation}
\item {\bf The singlet baryons}
\begin{equation}
	\begin{array}{rcccl}
	|(0,0),0,0,0 \rangle & : & \phi_A(\Lambda) & = & \frac{1}{\sqrt 6} [ |uds \rangle - |dus \rangle \\
	& & & & + |dsu \rangle - |sdu \rangle \\
	& & & & + |sud \rangle - |usd \rangle ]
	\end{array}
\end{equation}
\end{itemize}

\section{$^3P_0$ amplitudes: general expression}
\label{Ap1}
The color, spin and spatial parts of the $^{3}P_{0}$ amplitude $M_{A \rightarrow BC}(q_0) = \langle BC q_0 \ell J | T^\dag | A \rangle$, excluding the flavor couplings, were derived in a harmonic oscillator basis by Roberts and Silvestre-Brac (RSB) \cite{Roberts:1992}.
They did not consider a quark form factor ($\alpha_d = 0$) and used the following momenta
\begin{eqnarray}
\vec p_{\rho} &=& \frac{1}{2} (\vec p_1 - \vec p_2)  \mbox{ },
\nonumber\\
\vec p_{\lambda} &=& \frac{1}{3} (\vec p_1 + \vec p_2 - 2 \vec p_4)  \mbox{ },
\nonumber\\
\vec q_c &=& \frac{1}{2} (\vec p_3 - \vec p_5)  \mbox{ },
\nonumber\\
\vec q &=& \frac{1}{2} \left(\vec K_b - \vec K_c\right)  \mbox{ },
\nonumber\\
\vec P_{cm} &=& \vec K_b + \vec K_c  \mbox{ },
\label{momRSB}
\end{eqnarray}
with conjugate coordinates
\begin{eqnarray}
\vec \rho &=& \vec r_1 - \vec r_2 \mbox{ },
\nonumber\\
\vec \lambda &=& \frac{1}{2} (\vec r_1 + \vec r_2 - 2 \vec r_4)  \mbox{ }, 
\nonumber\\
\vec r_c &=& \vec r_3 - \vec r_5 \mbox{ }, 
\nonumber\\
\vec r &=& \frac{1}{3} \vec R_b - \frac{1}{2} \vec R_c \mbox{ }, 
\nonumber\\
\vec R_{cm} &=& \frac{1}{6} (\vec r_1 + \vec r_2 + \vec r_4) + \frac{1}{4} (\vec r_3 + \vec r_5) \mbox{ }.
\label{coordRSB}
\end{eqnarray} 
On the contrary, here we use the standard Jacobi coordinates and conjugate momenta for the initial baryon, $A$, and the final state baryon $B$ and meson $C$ \cite{Ferretti:PhD,SF:2015}, 
\begin{eqnarray}
\vec \rho &=& \frac{1}{\sqrt 2} (\vec r_1 - \vec r_2) \mbox{ },
\nonumber\\
\vec \lambda &=& \frac{1}{\sqrt 6} (\vec r_1 + \vec r_2 - 2 \vec r_4)  \mbox{ }, 
\nonumber\\
\vec r_c &=& \vec r_3 - \vec r_5 \mbox{ }, 
\nonumber\\
\vec r &=& \vec R_b - \vec R_c \mbox{ }, 
\nonumber\\
\vec R_{cm} &=& \frac{m_b \vec R_b + m_c \vec R_c}{m_b+m_c} \mbox{ }, 
\label{coord1}
\end{eqnarray}
where $m_i$ and $p_i$ are the mass and momentum of the quark $i$ (see Fig. \ref{fig:3P0decay}), $m_b = m_1 + m_2 + m_4$ and $m_c = m_3 + m_5$ the masses of the hadrons $B$ and $C$.  
Their center of mass coordinates are defined as
\begin{eqnarray}
\vec R_b &=& \frac{1}{3} (\vec r_1 + \vec r_2 + \vec r_4) \mbox{ }, 
\nonumber\\
\vec R_c &=& \frac{1}{2} (\vec r_3 + \vec r_5) \mbox{ },
\label{coord2} 
\end{eqnarray} 
The conjugate momenta are given by
\begin{eqnarray}
\vec p_{\rho} &=& \frac{1}{\sqrt 2} (\vec p_1 - \vec p_2)  \mbox{ },
\nonumber\\
\vec p_{\lambda} &=& \frac{1}{\sqrt 6} (\vec p_1 + \vec p_2 - 2 \vec p_4)  \mbox{ },
\nonumber\\
\vec q_c &=& \frac{1}{2} (\vec p_3 - \vec p_5)  \mbox{ },
\nonumber\\
\vec q &=& \frac{m_c \vec K_b - m_b \vec K_c}{m_b+m_c}  \mbox{ },
\nonumber\\
\vec P_{cm} &=& \vec K_b + \vec K_c  \mbox{ },
\label{coord3}
\end{eqnarray}
and
\begin{eqnarray}
\vec K_b &=& \vec p_1 + \vec p_2 + \vec p_4  \mbox{ },
\nonumber\\
\vec K_c &=& \vec p_3 + \vec p_5  \mbox{ }.
\label{coord4}
\end{eqnarray} 

The final result is \cite{Ferretti:PhD,SF:2015}
\begin{equation}
	\begin{array}{l}
	\label{eqn:Mabc}
	M_{A \rightarrow BC}(q_0) = 6 \gamma_0 \, \theta_{A \rightarrow BC} \, 
	\epsilon(l_{\lambda_b},l_c,L_{bc},l,l_{\lambda_a},L,q_0)  \mbox{ },
	\end{array}
\end{equation}
where the angular momenta of the baryons $A$ and $B$, $L_a$ and $L_b$, are the sum of the $\rho$ and $\lambda$ oscillator contributions, $l_\rho$ and $l_\lambda$, and the term $\theta_{A \rightarrow BC}$ contains the dependence on the color-spin-flavor part  
\begin{eqnarray}
\theta_{A \rightarrow BC} &=& \frac{\mathcal{F}_{A \rightarrow BC}}{3\sqrt{3}} (-1)^{l+l_{\lambda_a}} 
\sum_{L_{bc} S_{bc}} (-1)^{S_a - S_{bc} + L_{bc}} 
\nonumber\\
&& \left[ \begin{array}{ccc} J_{\rho} & \frac{1}{2} & S_b \\ \frac{1}{2} & \frac{1}{2} & S_c \\
S_a & 1 & S_{bc} \end{array} \right] \,  
\left[ \begin{array}{ccc} S_b & l_{\lambda_b} & J_b \\ S_c & l_c & J_c \\ S_{bc} & L_{bc} & J_{bc} \end{array} \right] \sum_L \hat{L}^2 
\nonumber\\
&& \left\{ \begin{array}{ccc} S_a & l_{\lambda_a} & J_a \\ L & S_{bc} & 1 \end{array} \right\} 
 \left\{ \begin{array}{ccc} S_{bc} & L_{bc} & J_{bc} \\ l & J_a & L \end{array} \right\} \mbox{ }. 
\end{eqnarray}
The coefficient 
$\mathcal{F}_{A \rightarrow BC}$ contains the flavor couplings and is defined as:
\begin{equation}
\mathcal{F}_{A \rightarrow BC} =\langle \phi_{B}\phi_{C} \lvert\phi_0\phi_{A}\rangle  \mbox{ }.
\end{equation}
Here, $\phi_0$ denotes the flavor wave function of the created quark-antiquark pair 
\begin{equation}
| \phi_0 \rangle = \frac{1}{\sqrt{2+(m_n/m_s)^2}} \Big[ | u\bar{u}\rangle + |d\bar{d}\rangle + \frac{m_n}{m_s} |s\bar{s}\rangle \Big] ~,
\label{new3p0}
\end{equation}
which, in the limit of equal quark masses, reduces to the usual expression for a flavor singlet. 
The coefficients in square brackets are proportional to 9-j coefficients 
\begin{equation}
\left[ \begin{array}{ccc} a & b & c \\ d & e & f \\ g & h & i \end{array} \right] =
\hat c \hat f \hat g \hat h \hat i \left\{ \begin{array}{ccc} a & b & c \\ d & e & f \\ g & h & i \end{array} \right\}  \mbox{ },
\end{equation}
where $\hat \ell = \sqrt{2 \ell + 1}$, 

Finally, $\epsilon(l_{\lambda_b},l_c,L_{bc},l,l_{\lambda_a},L,q_0)$ represents the spatial contribution \cite{Ferretti:PhD,SF:2015} 
\begin{widetext}
\begin{eqnarray}
\label{eqn:epsilon}
\epsilon(l_{\lambda_b},l_c,L_{bc},l,l_{\lambda_a},L,q_0) &=& \mathcal J 
\mathcal N_{n_{\lambda_a} l_{\lambda_a}}(\alpha_b) \, 
\mathcal N^*_{n_{\lambda_b} l_{\lambda_b}}(\alpha_b) \, \mathcal N^*_{n_c l_c}(\alpha_c)  (-1)^{L_{bc}} 
\frac{\exp(-F^2 q^2_0)}{2G^{l_{\lambda_a}+l_{\lambda_b}+l_c+4}} \sum_{l_1,l_2,l_3,l_4} 
C^{l_{\lambda_b}}_{l_1} \, C^{l_c}_{l_2}  
\nonumber\\ 
&& C^{1}_{l_3} \, C^{l_{\lambda_a}}_{l_4} \left( x -\sqrt{\frac{2}{3}} \right)^{l_1} 
\left( \frac{1}{2}-\sqrt{\frac{2}{3}}x \right)^{l_2} \left(-\sqrt{\frac{2}{3}}\right)^{l'_2} 
\left(\sqrt{\frac{2}{3}}x-1\right)^{l_3} \left(\sqrt{\frac{2}{3}}\right)^{l'_3} x^{l_4} 
\nonumber\\
&& \sum_{l_{12},l_5,l_6,l_7,l_8} (-1)^{l_{12} +l_6} \, \frac{\hat{l}_5}{\hat{L}} 
\left[ \begin{array}{ccc} l_1 & l'_1 & l_{\lambda_b} \\ l_2 & l'_2 & l_c \\ l_{12} & l_6 & L_{bc} \end{array} \right]
\left[ \begin{array}{ccc} l_3 & l'_3 & 1 \\ l_4 & l'_4 & l_{\lambda_a} \\ l_7 & l_8 & L \end{array} \right] 
\left\{ \begin{array}{ccc} l & l_{12} & l_5 \\ l_6 & L & L_{bc} \end{array} \right\} 
\nonumber\\
&& B^{l_{12}}_{l_1l_2} \, B^{l_{5}}_{ll_{12}} \, B^{l_6}_{l'_1l'_2} \, B^{l_7}_{l_3l_4} \, B^{l_8}_{l'_3l'_4} 
\sum_{\lambda , \mu,\nu} D_{\lambda \mu \nu}(x) \, I_\nu (l_5,l_6,l_7,l_8;L) 
\nonumber\\
&& \left( \frac{l'_1+l'_2+l'_3+l'_4+2 \mu +\nu +1}{2} \right)!
\frac{q^{l_1+l_2+l_3+l_4+2 \lambda +\nu}_0}{G^{2\mu +\nu-l_1-l_2-l_3-l_4}}  \mbox{ },
\end{eqnarray}
\end{widetext}
with
\begin{eqnarray}
C^{L}_{l} &=& \sqrt{\frac{4\pi(2L+1)!}{(2l+1)![2(L-l)+1]!}} 
\nonumber\\	
B^{l}_{l_1,l_2} &=& \frac{(-1)^l}{\sqrt{4\pi}}\hat{l_1}\hat{l_2}
\left(\begin{array}{ccc} l_1 & l_2 & l \\ 0 & 0 & 0 \end{array} \right)  \mbox{ }.
\end{eqnarray}
Here $l'_1=l_{\lambda_b}-l_1$, $l'_2=l_c-l_2$, $l'_3=1-l_3$, $l'_4=l_{\lambda_a}-l_4$ and $\hat l = \sqrt{2l+1}$.  
$\mathcal J=1/3\sqrt{3}$ is the Jacobian for the change of momenta $\{\vec p_1, \vec p_2, \vec p_3, \vec p_4, \vec p_5\}$ $\rightarrow$ $\{\vec p_{\rho}, \vec p_{\lambda}, \vec q_c , \vec q, \vec P_{cm} \}$, see Eqs.~(\ref{coord3},\ref{coord4}), $\mathcal N_{n_{\lambda_a} l_{\lambda_a}}(\alpha_b)$, $\mathcal N_{n_{\lambda_b} l_{\lambda_b}} (\alpha_b)$ and $\mathcal N_{n_c l_c}(\alpha_c)$, 
\begin{equation}
	\label{eqn:3DHO-rc}
	\begin{array}{rcl}
	\mathcal{N}_{n,L}(\alpha) = \sqrt{\frac{2 n !}{\Gamma(n+L+3/2)}} \mbox{ } \alpha^{-L-\frac{3}{2}} \mbox{ },
	\end{array}
\end{equation}
are the  normalization coefficients of the harmonic oscillator wave functions of the baryons $A$ and $B$,
\begin{widetext}
\begin{eqnarray}
	\label{eqn:3DHO-Plambda-Prho}
	\Phi_{n L M} (\vec p_\lambda,\vec p_\rho) & = & \sum_m \langle l_\rho, m; l_\lambda, M-m | L M\rangle \mathcal{N}_{n_\rho,l_\rho}(\alpha_b) 
	L_{n_\rho}^{l_\rho+1/2}(p_\rho^2/\alpha_b^2) \mbox{ } e^{-p_\rho^2/2\alpha_b^2} \mathcal Y_{l_\rho m} (\vec p_\rho) \nonumber\\ 
	& & \mathcal{N}_{n_\lambda,l_\lambda}(\alpha_b) L_{n_\lambda}^{l_\lambda+1/2}(p_\lambda^2/\alpha_b^2) \mbox{ } 
	e^{-p_\lambda^2/2\alpha_b^2} \mathcal Y_{l_\lambda M-m} (\vec p_\lambda)\mbox{ },
\end{eqnarray}
\end{widetext}
and meson $C$,
\begin{eqnarray}
	\label{eqn:3DHO-Pc}
	\Phi_{n_c l_c m_c} (\vec q_c) & = & \mathcal{N}_{n_c l_c}(\alpha_c) 
	L_{n_c}^{l_c+1/2}(q_c^2/\alpha_c^2) \mbox{ } e^{-q_c^2/2\alpha_c^2} \nonumber\\ 
	& & \mathcal Y_{l_c m_c} (\vec q_c) \mbox{ },
\end{eqnarray}
where $n$ is the number of nodes in the harmonic oscillator wave function, $L_{n}^{L+1/2}(\alpha p^2)$ a Laguerre polynomial and $\mathcal Y_{L M} (\vec p)$ a solid spherical harmonic.
The remaining coefficients are given by:
\begin{eqnarray}
G^2 &=& \alpha_b^2 + \frac{1}{3} \alpha_{\rm c}^2 + \frac{1}{3} \alpha_{\rm d}^2   \mbox{ },
\nonumber\\	
x &=& \frac{2\alpha_{\rm b}^2 + \alpha_{\rm c}^2 + 2\alpha_{\rm d}^2}{2\sqrt{6} \, G^2}  \mbox{ },
\nonumber\\	
F^2 &=& \frac{\alpha_{\rm b}^2 (12 \alpha_{\rm b}^2 + 5 \alpha_{\rm c}^2) + \alpha_{\rm d}^2 (20 \alpha_{\rm b}^2 + 3 \alpha_{\rm c}^2)}
{24 \left( 3 \alpha_{\rm b}^2 + \alpha_{\rm c}^2 + \alpha_{\rm d}^2 \right)}  \mbox{ }.
\label{eqn:G2xF2}
\end{eqnarray}
The present results for the $^{3}P_{0}$ amplitudes were obtained in a consistent way using the same Jacobi coordinates for the baryon wave functions and the $^{3}P_{0}$ matrix elements.
We observe that the coefficients of Eqs. (\ref{eqn:G2xF2}) also depend on the parameter $\alpha_{\rm d}$ of the Gaussian quark form factor, $V(\vec{p}_4 - \vec{p}_5) = e^{-\alpha_{\rm d}^2 (\vec p_4 - \vec p_5)^2/8}$.
The case $V = 1$ is a subcase of the previous one.

For the special case of ground-state baryons and pseudoscalar mesons, the orbital angular momenta vanish $l_{\lambda_a}=l_{\lambda_b}=l_c=L_{bc}=0$ and therefore $J_a=S_a$, $J_b=S_b$, $J_c=S_c=0$ and $J_{bc}=S_{bc}=J_b$. Due to the conservation of angular momentum and parity, the relative orbital angular momentum between the baryon $B$ and the meson $C$ is equal to $l=1=L$. 
As a result, the general expression for the $^{3}P_{0}$ transition amplitude simplifies considerably. 
The color-spin-flavor part $\theta_{A \rightarrow BC}$ reduces to
\begin{eqnarray}
\label{eqn:CSF-simplified}
\theta_{A \rightarrow BC} &=& -\frac{1}{3} \sqrt{\frac{2J_b+1}{2}} \, (-1)^{J_{\rho}+J_a-\frac{1}{2}}  \\
&& \times \left\{ \begin{array}{ccc} J_a & 1 & J_b \\ \frac{1}{2} & J_{\rho} & \frac{1}{2} \end{array} \right\} 
\mathcal{F}_{A \rightarrow BC}  \mbox{ },
\end{eqnarray}
where $J_{\rho}$ the total angular momentum of the first two quarks and $\mathcal{F}_{A \rightarrow BC}$ the flavor matrix elements of the $A \rightarrow BC$ transition \cite{Roberts:1992}. The spatial contribution simplifies to 
\begin{eqnarray}
\epsilon(q_0) = - \frac{1}{2} \left( \frac{9 \alpha_b^4 \alpha_c^2}{\pi} \right)^{3/4}
\frac{(4 \alpha_b^2 + \alpha_c^2) \, q_0 \mbox{e}^{-F^2 q_0^2}}{\left( 3 \alpha_b^2 + \alpha_c^2 + \alpha_d^2 \right)^{5/2}} 
 \mbox{ }.
\end{eqnarray}

\section{Flavor couplings}
\label{Ap2}

In the following, we give the flavor coefficients ${\cal F}_{A \rightarrow BC}$. These expressions are valid for both 
pseudoscalar and vector mesons. 

\begin{itemize}
\item {\bf $A_8 \rightarrow B_8 + C_8$ couplings.}

For octet baryons the flavor wave function has two components, labeled by $\rho$ and $\lambda$, both of which 
give rise to the flavor coupling coefficient 
\begin{eqnarray}
\mathcal{F}_{A \rightarrow BC}^{\rho} &=& \langle \phi_{\rho}(B) \phi(C) \lvert\phi_0\phi_{\rho}(A)\rangle ~,
\nonumber\\
\mathcal{F}_{A \rightarrow BC}^{\lambda} &=& \langle \phi_{\lambda}(B) \phi(C) \lvert\phi_0\phi_{\lambda}(A)\rangle ~, 
\end{eqnarray}
where we introduced the superscripts $\rho$ and $\lambda$ to distinguish the two contributions. 
In this case, the flavor couplings are given by
\begin{eqnarray}
\left( \begin{array}{c} N \\ \Sigma \\ \Lambda \\ \Xi \end{array} \right) \rightarrow 
\left( \begin{array}{ccccc} N \pi & N \eta_8 & \Sigma K & \Lambda K & \\
N \bar{K} & \Sigma \pi & \Lambda \pi & \Sigma \eta_8 & \Xi K \\
N \bar{K} & \Sigma \pi & \Lambda \eta_8 & \Xi K & \\
\Sigma \bar{K} & \Lambda \bar{K} & \Xi \pi & \Xi \eta_8 & \end{array} \right) = \nonumber \\
{\cal N} \left( \begin{array}{ccccc} 
\frac{1}{\sqrt{2}} & \frac{1}{3\sqrt{2}} & 0 & -\frac{\sqrt{2}}{3} \frac{m_n}{m_s} & \\ \\
0 & \frac{1}{\sqrt{3}} & \frac{1}{3\sqrt{2}} & \frac{1}{3\sqrt{2}} & -\frac{1}{\sqrt{3}}\frac{m_n}{m_s}  \\ \\
\frac{2}{3} & -\frac{1}{\sqrt 6} & \frac{1-4\frac{m_n}{m_s}}{9\sqrt{2}} & \frac{1}{3} \frac{m_n}{m_s} & \\ \\
\frac{1}{\sqrt{2}} & \frac{1}{3\sqrt{2}} & 0 & -\frac{\sqrt{2}}{3} \frac{m_n}{m_s} & \end{array} \right)
\label{eqn:888a}
\end{eqnarray}
for the $\rho$ component and by 
\begin{eqnarray}
{\cal N} \left( \begin{array}{ccccc} 
-\frac{1}{3\sqrt{2}} & \frac{1}{3\sqrt{2}} & \frac{\sqrt{2}}{3} \frac{m_n}{m_s} & 0 & \\ \\
\frac{2}{3\sqrt{3}} & \frac{1}{3\sqrt{3}} & -\frac{1}{3\sqrt{2}} & \frac{1-4\frac{m_n}{m_s}}{9\sqrt{2}} & \frac{1}{3\sqrt{3}}\frac{m_n}{m_s}  \\ \\
0 & \frac{1}{\sqrt 6} & \frac{1}{3\sqrt{2}} & \frac{1}{3} \frac{m_n}{m_s} & \\ \\
-\frac{1}{3\sqrt{2}} & \frac{1}{3\sqrt{2}} & \frac{\sqrt{2}}{3} & \frac{\sqrt{2}(1-\frac{m_n}{m_s})}{9} & \end{array} \right)
\label{eqn:888b}
\end{eqnarray}
for the $\lambda$ component. 
The normalization coefficient is given by 
\begin{equation}
{\cal N} = \sqrt{\frac{3}{{2+\left(\frac{m_n}{m_s}\right)^2}}} ~. 
\end{equation}
 
\item {\bf $A_8 \rightarrow B_8 + C_1$ couplings.}

The flavor coefficients for octet baryons in combination with a singlet meson 
are given by 
\begin{eqnarray}
\left( \begin{array}{c} N \\ \\ \Sigma \\  \\ \Lambda \\ \\\Xi \end{array} \right) \rightarrow 
\left( \begin{array}{c} N \eta_1 \\ \\ \Sigma \eta_1 \\ \\ \Lambda \eta_1 \\ \\ \Xi \eta_1 \end{array} \right) = 
{\cal N} \left( \begin{array}{c} \frac{1}{3} \\ \\ \frac{1}{3} \\ \\ \frac{1+2\frac{m_n}{m_s}}{9} \\ \\ \frac{1}{3} \frac{m_n}{m_s} \end{array} \right) 
\end{eqnarray}
for the $\rho$ component and by 
\begin{eqnarray}
{\cal N} \left( \begin{array}{c} \frac{1}{3} \\ \\ \frac{1+2\frac{m_n}{m_s}}{9} \\ \\ \frac{1}{3} \\ \\ \frac{2+\frac{m_n}{m_s}}{9} \end{array} \right) 
\end{eqnarray}
for the $\lambda$ component. 

\item {\bf $A_8 \rightarrow B_{10} + C_8$ couplings.} 

In this case the final baryon belongs to the decuplet 
which only couples to the $\lambda$ component of the initial octet baryon
\begin{equation}
\mathcal{F}_{A \rightarrow BC} = \langle \phi_{S}(B) \phi(C) \lvert\phi_0\phi_{\lambda}(A)\rangle  \mbox{ }.
\end{equation}
The flavor coefficients are given by
\begin{eqnarray}
\left( \begin{array}{c} N \\ \Sigma \\ \Lambda \\ \Xi \end{array} \right) \rightarrow 
\left( \begin{array}{cccc} & \Delta \pi & \Sigma^* K & \\
\Delta \bar{K} & \Sigma^* \pi & \Sigma^* \eta_8 & \Xi^* K \\
& \Sigma^* \pi & \Xi^* K & \\
\Sigma^* \bar{K} & \Xi^* \pi & \Xi^* \eta_8 & \Omega K \end{array} \right) = 
\nonumber\\
{\cal N} \left( \begin{array}{cccc} & \frac{2}{3} & -\frac{1}{3}\frac{m_n}{m_s} & \\  \\
-\frac{2 \sqrt 2}{3 \sqrt 3} & \frac{\sqrt 2}{3 \sqrt 3} & \frac{1+2\frac{m_n}{m_s} }{9} & -\frac{\sqrt 2}{3\sqrt 3} \frac{m_n}{m_s}\\ \\
& \frac{1}{\sqrt 3} & -\frac{\sqrt 2}{3}\frac{m_n}{m_s} & \\  \\
-\frac{1}{3} & \frac{1}{3} & \frac{1+2\frac{m_n}{m_s} }{9} & -\frac{\sqrt 2}{3}\frac{m_n}{m_s} \end{array} \right) 
\end{eqnarray}

\item {\bf $A_{10} \rightarrow B_8 + C_8$ couplings.}

As in the previous case, the decuplet baryon  
only couples to the $\lambda$ component of the octet baryon
\begin{equation}
\mathcal{F}_{A \rightarrow BC} = \langle \phi_{\lambda}(B) \phi(C) \lvert\phi_0\phi_{S}(A)\rangle  \mbox{ }.
\end{equation}
The flavor coefficients are given by
\begin{eqnarray}
\left( \begin{array}{c} \Delta \\ \Sigma^* \\ \Xi^* \\ \Omega \end{array} \right) \rightarrow 
\left( \begin{array}{ccccc} & N \pi & \Sigma K & & \\
N \bar{K} & \Sigma \pi & \Lambda \pi & \Sigma \eta_8 & \Xi K \\
\Sigma \bar{K} & \Lambda \bar{K} & \Xi \pi & \Xi \eta_8 & \\
& & \Xi \bar{K} & & \\\end{array} \right) = 
\nonumber\\ 
{\cal N} \left( \begin{array}{ccccc} & -\frac{\sqrt{2}}{3} & \frac{\sqrt{2}}{3}\frac{m_n}{m_s}  & & \\ \\
-\frac{\sqrt{2}}{3\sqrt 3} & \frac{\sqrt{2}}{3\sqrt 3} & -\frac{1}{3} & \frac{1+2\frac{m_n}{m_s} }{9} & \frac{\sqrt{2}}{3\sqrt 3} \frac{m_n}{m_s} \\  \\
\frac{1}{3} & -\frac{1}{3} & \frac{1}{3} & \frac{1+2\frac{m_n}{m_s} }{9}  & \\ \\
& & \frac{2}{9} & & \end{array} \right) 
\end{eqnarray}

\item {\bf $A_{10} \rightarrow B_{10} + C_8$ couplings.} 

The flavor coefficients for decuplet baryons in combination with a octet meson 
\begin{equation}
\mathcal{F}_{A \rightarrow BC} = \langle \phi_{S}(B) \phi(C) \lvert\phi_0\phi_{S}(A)\rangle  \mbox{ }.
\end{equation}
are given by 
\begin{eqnarray}
\left( \begin{array}{c} \Delta \\ \Sigma^* \\ \Xi^* \\ \Omega \end{array} \right) \rightarrow 
\left( \begin{array}{cccc} \Delta \pi & \Delta \eta_8 & \Sigma^* K & \\
\Delta \bar{K} & \Sigma^* \pi & \Sigma^* \eta_8 & \Xi^* K \\
\Sigma^* \bar{K} & \Xi^* \pi & \Xi^* \eta_8 & \Omega K \\
& \Xi^* \bar{K} & \Omega \eta_8 & \end{array} \right) = 
\nonumber\\
{\cal N} \left( \begin{array}{cccc} 
\frac{\sqrt{5}}{3\sqrt{2}} & \frac{1}{3\sqrt{2}} & -\frac{1}{3}\frac{m_n}{m_s}  & \\ \\
\frac{2}{3\sqrt 3} & \frac{2}{3\sqrt 3} & -\frac{\sqrt{2}(\frac{m_n}{m_s}-1)}{9} & -\frac{2}{3\sqrt 3}\frac{m_n}{m_s}  \\  \\
\frac{\sqrt{2}}{3} & \frac{1}{3\sqrt 2} & -\frac{(4\frac{m_n}{m_s}-1)}{9\sqrt{2}} & -\frac{1}{3} \frac{m_n}{m_s} \\  \\
& \frac{\sqrt{2}}{3} & -\frac{\sqrt{2}}{3}\frac{m_n}{m_s}  & \end{array} \right) 
\end{eqnarray}

\item {\bf $A_{10} \rightarrow B_{10} + C_1$ couplings.}

The flavor coefficients for decuplet baryons in combination with a singlet meson 
are given by 
\begin{eqnarray}
\left( \begin{array}{c} \Delta \\ \\ \Sigma^* \\ \\ \Xi^* \\ \\ \Omega \end{array} \right) \rightarrow 
\left( \begin{array}{c} \Delta \eta_1 \\ \\ \Sigma^* \eta_1 \\ \\ \Xi^* \eta_1 \\ \\  \Omega \eta_1 \end{array} \right) = 
{\cal N} \left( \begin{array}{c} \frac{1}{3} \\ \\ \frac{2+\frac{m_n}{m_s}}{9} \\  \\ \frac{1+2\frac{m_n}{m_s}}{9}\\  \\ \frac{1}{3} \frac{m_n}{m_s}  \end{array} \right) 
\label{eqn:10101}
\end{eqnarray}
\end{itemize}

\section{Comparison between Elementary Meson Emission and $^3P_0$ Models}
\label{EME vs 3P0}
In the following, we shall compare the Elementary Emission (EME) and $^3P_0$ Models.
In the EME model,  the decay proceeds via the emission of a meson in terms of an elementary quantum.  On the contrary, in the $^3P_0$ Model the decay process is described in terms of the creation of an additional quark-antiquark pair. 

In order to do a comparison between EME and $^3P_0$  models, we consider the simplest form for the EME transition operator \cite{HadronT},   \begin{eqnarray}
{\cal H}_s = \frac{g}{(2\pi)^{3/2} (2k_0)^{1/2}} 
 X^c\left[
\, (\vec{\sigma} \cdot \vec{k}) \mbox{e}^{-i \vec{k} \cdot \vec{r}} \right. \\
\left. + \frac{k_0}{2m}\ \, \vec{\sigma}\cdot
(\vec{p} \, \mbox{e}^{-i \vec{k} \cdot \vec{r}} +
\mbox{e}^{-i \vec{k} \cdot \vec{r}} \, \vec{p}) \right] ~,  
\label{hs}
\end{eqnarray}
where $g=G_{\rm cqq}/2m$ is the ratio between the meson-emission strength (emission of a meson $C$ by a quark) and the quark mass, $X^c$ is the flavor
operator related to the emission of meson $C$. 
On the other hand, the $^3P_0$ operator in matrix form is 
\begin{eqnarray}
( \langle C |T | 0\rangle )_{i'_3i_3}= \sum_{i_5} \Psi^*_{Ci_3i_5}P_{i'_3i_5}=PC^\dagger\label{3P0M}  \mbox{ },
\end{eqnarray}
where $T$ is the $^3P_0$ operator 
\begin{equation}
\begin{array}{rcl}
	T^{\dagger} & = & -3 \, \int d \vec{p}_4 \, d \vec{p}_5 \, \delta(\vec{p}_4 + \vec{p}_5) \, C_{45} F_{45} \left[ \chi_{45} \right. \\
	& & \left. \times \, {\cal Y}_{1}(\vec{p}_4 - \vec{p}_5) \right]^{(0)}_0  b_4^{\dagger}(\vec{p}_4) \, d_5^{\dagger}(\vec{p}_5)    \mbox{ }.
\label{3p0-T02}
\end{array}
\end{equation}
In Ref. \cite{HadronT}, it is shown that equation (\ref{3P0M}) can also be written as
\begin{equation}
\begin{array}{l}
PC^\dagger=\frac{\gamma}{2(6\pi)^{1/2}}e^{-i \vec{k}_c \cdot \vec{r}}X^c\left [ \vec{\sigma}\cdot(\vec{k}_c-\vec{p})\right] \psi_C(2\vec{p}-\vec{k}_c)
\end{array}  \mbox{ }.
\label{hs3}
\end{equation}

If we compare Eqs. (\ref{hs}) and (\ref{hs3}), we can observe that in the $^3P_0$ model there are direct and recoil terms with equal weights; on the contrary,   the recoil term in the EME has a different weight factor $k_0/2m$. 
To have a perfect correspondence between EME and $^3P_0$ models, this factor in the recoil term has to be taken equal to $k_0/2m=1$. 
However, this factor has been taken as a free parameter in BIL \cite{Bijker:1996tr,Bijker:1994yr} (and also in several other studies, like \cite{Koniuk:1979vy,Bijker:1996tr,Bijker:1994yr}) and, by fitting, it turned out to be equal to 0.04, which means that the recoil term is practically absent in BIL. 

It is worthwhile noting that in the calculation of open-flavor strong decays for $L_b=L_c=0$, the final $BC$ state is characterized by a relative angular momentum $\ell=L_a\pm1$, where $L_a$ is the angular momentum of the initial state, $A$. In Ref. \cite{HadronT}, it is shown that in the EME, the partial wave $\ell=L_a+1$ is essentially given by the direct term with a small correction from the recoil one, while $\ell=L_a-1$ receives its essential contribution from the recoil term. Thus, the presence of the factor $k_0/2m$ in the EME operator can give rise to different results with respect to the $^3P_0$ model. 

Another important difference is due to the wave function factor $\psi_C(2\vec{p}-\vec{k}_c)$ in Eq. (\ref{hs3}), related to the composite structure of the meson $C$, which gives rise to a non local character of the operator.

\begin{acknowledgments}
This work was supported in part by INFN, EPOS, Italy and by PAPIIT-DGAPA, Mexico (Grant No. IN107314).
\end{acknowledgments}

\end{document}